\documentclass{article}

\usepackage{PRIMEarxiv}

\usepackage[utf8]{inputenc} 
\usepackage[T1]{fontenc}    
\usepackage{hyperref}       
\usepackage{url}            
\usepackage{booktabs}       
\usepackage{amsfonts}       
\usepackage{nicefrac}       
\usepackage{microtype}      
\usepackage{lipsum}
\usepackage{fancyhdr}       
\usepackage{graphicx}       
\graphicspath{{media/}}     

\usepackage{natbib}
\usepackage{amsthm}
\usepackage{graphics, caption}
\usepackage{bm}
\usepackage[plain,noend]{algorithm2e}
\usepackage{xcolor}
\usepackage{booktabs, makecell, multirow, tabularx}     
\usepackage{mathtools} 
\usepackage{verbatim}

\newtheorem{theorem}{Theorem}
\newcommand{\RR}{I\!\!R} 
\newcommand{\bigo}[1]{\mathcal{O}\left( #1 \right)}

\title{Information Borrowing in Regression Models
\thanks{\textit{This work was supported by the NIH/NIAID under grant 5-R01-AI136664}} 
}

\author{
  Amy Zhang, Le Bao \\
  Department of Statistics, The Pennsylvania State University \\
  \texttt{lebao@psu.edu} \\
   \And
  Michael J. Daniels \\
  Department of Statistics, University of Florida\\
}

\begin{document}
\maketitle

\begin{abstract}
Model development often takes data structure, subject matter considerations, model assumptions, and goodness of fit into consideration. To diagnose issues with any of these factors, it can be helpful to understand regression model estimates at a more granular level. We propose a new method for decomposing point estimates from a regression model via weights placed on data clusters. The weights are informed only by the model specification and data availability and thus can be used to explicitly link the effects of data imbalance and model assumptions to actual model estimates. The weight matrix has been understood in linear models as the hat matrix in the existing literature. We extend it to Bayesian hierarchical regression models that incorporate prior information and complicated dependence structures through the covariance among random effects. We show that the model weights, which we call borrowing factors, generalize shrinkage and information borrowing to all regression models. In contrast, the focus of the hat matrix has been mainly on the diagonal elements indicating the amount of leverage. We also provide metrics that summarize the borrowing factors and are practically useful. We present the theoretical properties of the borrowing factors and associated metrics and demonstrate their usage in two examples. By explicitly quantifying borrowing and shrinkage, researchers can better incorporate domain knowledge and evaluate model performance and the impacts of data properties such as data imbalance or influential points. 
\end{abstract}

\keywords{Information borrowing \and Regression \and Bayesian hierarchical model}

\section{Introduction}\label{ssbf-sec:intro}

Model development is often an iterative process, particularly in challenging settings with high-dimensional feature sets or complex dependency structures. Data properties, subject matter considerations, model assumptions, and goodness of fit are all factors that are taken into consideration, and multiple models may be evaluated and compared to each other. To diagnose issues with any of these factors, it can be helpful to understand regression model estimates at a more granular level. We propose to understand regression model estimates by expressing them as a function of a vector of weights placed on each data point. It offers the intuitive interpretation that estimates are formed by ``borrowing'' information from other data points, with the weight being the amount borrowed. As such, we call the weights ``borrowing factors''. 

This granular decomposition of regression model estimates can be particularly helpful for Bayesian hierarchical regression models, where a shared hyperprior is placed on model parameters to pool information between them and improve model estimates \citep{gelman2007data}. Information pooling has historically been understood through the lens of the James-Stein estimator. Given observed data $Y_i \sim N(\alpha_i, \phi^2)$, $j = 1, \dots, J$, \citet{stein1956inadmissibility} developed a biased estimator which improves upon the unbiased ordinary least squares (OLS) estimator for $\alpha \in \RR^P$, $P \ge 3$, under squared loss. This result was later improved by \citet{james1992estimation} and dubbed the ``James-Stein estimator''. Given data $Y_j \sim N(\alpha_j, 1)$, the James-Stein estimator is 
$$\hat{\alpha}^{\text{JS}}_j = \mu_j + \frac{1 - P - 2}{S}(Y_j - \mu_j), \quad S = \sum (Y_j - \mu_j)^2,$$ 
where $\mu_j$ is an initial guess at $\alpha_j$; James and Stein used the global data mean $\mu = \bar{Y}$. \citet{efron1973stein} showed the James-Stein estimator is one of a class of empirical Bayesian methods that dominate the OLS estimator under squared loss by shrinking estimates for $\alpha$ towards some global mean $\mu$, producing biased estimates but reducing the variance of the estimator, resulting in a lower overall loss. This shrinkage towards the mean is referred to as information pooling.

To our best knowledge, information pooling has been only quantified for simple one-way models where $Y_i \sim N(\alpha_i, \phi_i^2)$, $\alpha_i \sim N(a_0, \sigma^2)$. Assuming $a_0, \phi,$ and $\sigma$ are known, some algebra and simplification results in the empirical Bayes estimator
$$\hat{Y}_i = \lambda a_0 + (1 - \lambda)\bar{Y}_i, \quad \lambda \in [0, 1],$$ 
where $\lambda$ was called the ``pooling factor'' by \citet{gelman2006bayesian}. Bayesian hierarchical models have been shown to perform well in several empirical studies \citep{morris1983parametric, gelman2007data}, and information pooling is often cited as the reason. However, information pooling has not been explicitly quantified in scenarios outside of the one-way setting, which has limited its use in applications; one of few examples is \citet{gelman2006bayesian}. Our method quantifies information pooling for any regression model and can identify patterns of information borrowing; for example, assessing whether the information is pooled evenly or unevenly. We can then confirm whether model estimates are in accordance with domain knowledge, which is often the deciding factor between models that perform similarly well based on the goodness of fit.

Explicitly quantifying information pooling can be particularly useful when the data are highly imbalanced, which can lead to biased estimates \citep{gelman2007data}. In many applications, this can result in different decisions being made. So the effects of data imbalance are often evaluated through extensive simulation studies, some recent examples of which include \citet{eager2017mixed}, \citet{mccarron2011bayesian}, and \citet{thabtah2020data}. By linking the data availability to the degree of information pooling, we can directly quantify the impact of data imbalance on the model estimates without simulation. Note that the simulation can be difficult when there are many sources of potential data imbalance to quantify and examine. It can also be challenging to translate conclusions from simulations to a specific observed data set. 

The weight matrix we consider here is the hat matrix in linear models;
the focus of the hat matrix has been mainly on the diagonal elements indicating the amount of leverage. Our proposed method uses the weight matrix to quantify the impact of influential observations on point estimates in Bayesian hierarchical regression models. We also introduce a metric to identify point estimates that rely heavily on a specific subset of data, called sum squares of borrowing factors (SSBF). After identifying influential observations in a model, researchers may exclude them from the final analysis  \citep{Belsley2005regression,Chatterjee2009sensitivity}. However, the decision is typically made using a combination of domain knowledge and influence analysis metrics. The borrowing factors can help such decisions by further identifying which point estimates are impacted the most by high-leverage observations and to what degree. If an observation is highly influential, but its influence is mostly limited to a small and specific subset of related observations, subject matter and model considerations can then inform whether to remove or include the observation.

In Section~\ref{ssbf-sec:notation}, we formally define the borrowing factors and introduce the sum squares of borrowing factors (SSBF), which is a summary of the information borrowing pattern for each point, as well as some useful terminology. In Section~\ref{ssbf-sec:theorems}, we describe theoretical properties of the borrowing factors and SSBF. We show that the borrowing factors are connected to the pooling factor and demonstrate SSBF's connection to two influence analysis metrics. In the next two sections, we illustrate how the borrowing factors and SSBF can link the effects of model assumptions and data availability to model estimates using two example data sets. Section~\ref{ssbf-sec:radon} shows how we can explicitly quantify the effects of data imbalance using the Radon data set \citep{gelman2007data}. Section~\ref{ssbf-sec:srd} uses the Scottish respiratory disease (SRD) data to show how model assumptions can be linked to model estimates and how the borrowing factors and SSBF can be used to provide context to influence analysis and quantify the impact of influential points on model estimates. We offer discussions in Section~\ref{ssbf-sec:discussion}.

\section{Quantifying shrinkage and information borrowing}\label{ssbf-sec:notation}
In this section, we provide an overview of our approach, with detailed discussion of theoretical properties in Section~\ref{ssbf-sec:theorems}. We discuss the Bayesian setting first.  Let $\bm{Y} \in \RR^N$ denote a continuous response vector that follows
\begin{equation}
\begin{split}
 \bm{Y}| \bm{\beta}, \Phi \sim N(X_1\bm{\beta}_1 + X_2\bm{\beta}_2, \Phi), \\
 \bm{\beta}_1 \sim N(\alpha_1, C), \quad \bm{\beta}_2 | \Sigma \sim N(\alpha_2, \Sigma), \\
 \Sigma  \sim f(\Sigma), \quad \Phi \sim f(\Phi),
\end{split}
\label{ssbf-eqn:normlinreg}
\end{equation}
where $X \coloneqq \begin{bmatrix}X_1 & X_2\end{bmatrix} \in \RR^{N\times P}$ is the design matrix, $\bm{\beta}_1 \in \RR^{P_1}$, $\bm{\beta}_2 \in \RR^{P_2}$ s.t. $\bm{\beta} \coloneqq \begin{bmatrix}\bm{\beta}_1' & \bm{\beta}_2'\end{bmatrix}' \in \RR^{P}$, $C \in \RR^{P_1 \times P_1}$ is positive-definite and typically a diagonal matrix, $\Sigma \in \RR^{P_2 \times P_2}$ is positive-definite, and $\Phi \in \RR^{N \times N}$ is diagonal and positive-definite. We assume  $\mathbf{1} \in \text{span}(X_1)$, where $\mathbf{1}$ is the $N$-length vector of ones, which is satisfied when the fixed effects include a global intercept or set of intercepts which partition the data. We take $\alpha_1 = \alpha_2 = 0$ throughout this paper, without loss of generality. $C$ is treated as fixed, often with large variances, and thus $\bm{\beta}_1$ are referred to as the fixed effects.  Random variance hyperparameters such as $\Sigma$ reflect the dependency among the $\bm{\beta}_2$; the effect is to pool information among related units and shrink them towards a common mean, thus the $\bm{\beta}_2$ are referred to as random effects. $\Sigma$ can take many forms, as long as it is positive-definite.

When modeling data as in (\ref{ssbf-eqn:normlinreg}), the posterior mean for $X\bm{\beta}$ conditioned on variance parameters has the form
\begin{equation}
\label{ssbf-eqn:axe_linregr}
   E[X\bm{\beta} | \Sigma, \Phi, \bm{Y}] \sim  XVX'\Phi^{-1}\bm{Y},  \quad V = \left(X'\Phi^{-1}X + \begin{bmatrix}C^{-1} & 0 \\ 0 & \Sigma^{-1}\end{bmatrix}\right)^{-1},
\end{equation}
where $C^{-1}$ is taken as the matrix of $0$s, which {corresponds to the assumption that the fixed effects have infinite variance}. \citet{kass1989approximate} show that the posterior mean $E[\bm{\beta} | \bm{Y}] = E[\bm{\beta} | \bm{Y}, \hat{\Sigma}_{\text{EB}}](1 + \bigo{P_2^{-1}})$, where $\hat{\Sigma}_{\text{EB}}$ denotes the Empirical Bayes estimates and $\hat{\Sigma}_{\text{EB}}$ in turn approximates posterior mean $\hat{\Sigma} = E[\Sigma | \bm{Y}]$ with order $\bigo{P_2^{-1}}$. Conditioning on variance parameters and using the posterior means $\hat{\Sigma}$ and $\hat{\Phi}$ as plug-in estimates in (\ref{ssbf-eqn:axe_linregr}) then produces estimates which approximate the posterior mean. The accuracy of this approximation is simple to determine by comparing the conditional expectation $E[X\bm{\beta}| \hat{\Sigma}, \hat{\Phi}, \bm{Y}]$ to the posterior expectation $E[X\bm{\beta} | \bm{Y}]$.

In the frequentist setting, the coefficients $\bm{\bm{\beta}_1}$ and variance parameters $\Phi$ and $\Sigma$ are non-random and fixed at their estimated values, with
\begin{align*}
    {\bm{Y}} &= X_1\bm{\beta}_1 + X_2\bm{\beta}_2 + {\bm{\epsilon}}, \\
    &\bm{\epsilon} \sim N(0, \Phi), \quad \bm{\beta}_2 \sim N(0, \Sigma).
\end{align*}
So, (\ref{ssbf-eqn:axe_linregr}) directly expresses the fitted values for the frequentist regression model and is not an approximation. 

In the case of a generalized linear model, we approximate the non-linear data-level model $f(Y | \beta)$ with a normal distribution having the same moments. This was shown by \citet{daniels1998note} to be a Laplace approximation with the same asymptotic error. The accuracy of the approximate is straightforward to determine by numerically comparing $E[X\bm{\beta} | \bm{Y}]$ to its normal approximate.

Equation (\ref{ssbf-eqn:axe_linregr}) expresses mean estimates $\hat{Y}_i$, $i = 1, \dots, N$, as a weighted average of the response data $\bm{Y}$, where the $N \times N$ matrix of weights is
\begin{equation}\label{ssbf-eqn:w}
    W \coloneqq XVX'\Phi^{-1}
\end{equation}
and is informed only by the model specification and data availability, not the response. How data availability and model specification impact model estimates can then be wholly determined by examining $W$, and an entry $w_{ij}$ in the $i^{th}$ row and $j^{th}$ column of $W$ can be thought of as the amount of information borrowed from $Y_j$  for point estimate $\hat{Y}_i$. This allows us to explicitly quantify the amount of information borrowing for all model estimates. 

How to interpret $W$ such that we can clearly link data availability or model assumptions to model estimates? We aggregate over $w_{ij}$'s to determine the amount borrowed from a set of points $J \subset \{1, \dots, N\}$. We refer to both $w_{ij}$ and $\sum_{j \in J}w_{ij}$ as ``borrowing factors'', with the latter denoted as $b_{iJ}$. The borrowing factors can then be linked to data availability, model covariates, or other quantities of interest. This can help to identify higher-level patterns of information borrowing and determine which lenders are the most impactful for any specific point estimate $\hat{Y}_i$. After understanding how model assumptions and the data availability lead to point estimates, researchers can verify whether model estimates are generated in ways aligned with subject matter considerations. For instance, in a model of standardized test scores with school, class, and age as covariates, the borrowing factors can determine whether the estimated standardized test score of a student borrows more from students of the same school, students of the same class, or students of the same age group (younger v.s. older).

When $i = j$, $w_{ij}$ is the amount of information borrowed from a point estimate's own data. It is helpful to separately consider such cases---let $x_i'$ denote the $i^{th}$ row of the model matrix $X$, and let $B_i=\{j \in 1, \dots, N: x_j = x_i, \phi_j = \phi_{i}\}$ {indicate rows that have the identical design covariates and variance with the $i^{th}$ row}, where $\phi_i^2$ is the $i^{th}$ diagonal entry of $\Phi$. We denote the cardinality of $B_i$ as $n_i$. Note that $w_{ij} = w_{ii}$ for all $j \in B_i$, thus any of the $Y_{B_i}$ can be exchanged with each other and obtain the same model estimates. We call the set of indices ${B_i}$ the borrowers or the borrower cluster. The shrinkage factor is the total weight placed on the borrower cluster,
\begin{equation}
    \label{ssbf-eqn:sf}
    b_{iB_i} = \sum_{j \in B_i}w_{ij}.
\end{equation}
All other points are referred to as the lenders, $L_i = \{j \in 1, \dots, N: x_j \neq x_i\}$.  The {pooling factor} is the total weight placed on lenders,
\begin{equation}
    \label{ssbf-eqn:pf}
    b_{iL_i} = 1 - b_{iB_i} = b_{iL_i}.
\end{equation} 
If a point estimate $\hat{Y}_i$ has lower pooling factor, then its value will be closer to $\bar{Y}_i$. 

The terms shrinkage and pooling factors originate from the Bayesian literature for simple one-way models, $Y_i \sim N(\alpha_i, \phi_i^2)$, $\alpha_i \sim N(a_0, \sigma^2)$ \citep{efron1975data, gelman2006bayesian} and the definitions we present here extend the definition to all regression models, as we show in Section~\ref{ssbf-sec:thms_bf}. They help to summarize how similar a point estimate $\hat{Y}_i$ is to its data mean $\bar{Y}_i$ versus how much is borrowed, which by itself can be helpful for understanding model estimates. However, they do not contain information on which lenders are borrowed from the most and thus cannot explain what higher-level patterns of information borrowing exist. We may have some intuition; for example, if the data are imbalanced, we may presume that those clusters with less data will borrow more from other clusters, and for that borrowing to come largely from clusters with more data, but this has not been explicitly quantified for any model in the literature. 

We also {propose a metric that summarizes} total borrowing in each row of $W$, the sum squares of borrowing factors (SSBF), where
\begin{equation}
    \label{ssbf-eqn:ssbf}
    \text{SSBF}_i = \sum_{j \in L_i}w_{ij}^2.
\end{equation}
SSBF is similar to the pooling factor in that it aggregates over the borrowing factors of the lenders but it uses their squared values. Point estimates will thus have higher SSBF if they place high individual weight on lenders and low SSBF if no lender has particularly large weight; in fact, we show later in (\ref{ssbf-eqn:ssbfvar}) that SSBF is proportional to the sample variance of borrowing factors. Thus points with high SSBF have more distinct borrowing patterns, with some lenders having high individual borrowing factors, based on a relationship they share with the borrower cluster. {Understanding how SSBF changes with data availability, model covariates, or other metrics of interest can help identify  borrowing patterns.} SSBF is also related to both the retrospective value of sample information \citep{parsons2018value} and \citet{pena2005new}'s metric $S_i$ in the influence analysis literature and can be thought of as the total influence of all lenders due solely to the data availability. In some scenarios, it can also be interpreted as model uncertainty for estimate $\hat{Y}_i$. We show these properties and discuss them in more detail in Section~\ref{ssbf-sec:thms_ssbf}. 

To identify borrowing patterns for a borrower cluster $B_i$, it is often helpful to partition the lenders $L_i$ into a set of relationship groups, where the groups are determined based on the lenders' similarity to the borrower. For models with clustered data, a good starting point to define relationship groups is to examine the locations of non-zero entries of $x_ix_j'$ and to group together those points $j$ that have the same non-zero locations. Zero values of $x_i$ indicate that the corresponding entry in $\bm{\beta}$ does not contribute to $\hat{Y}_i$; the non-zero entries in $x_ix_j'$ then correspond to coefficients which contribute to both $\hat{Y}_i$ and $\hat{Y}_j$. For example, given a nested model with $E[{Y}_{ljk} | a_0, \alpha_j, \alpha_{jk}] = a_0 + \alpha_j + \alpha_{jk}$, where ${Y}_{ljk}$ is the standardized test score of student $l$ from school $k$ of school district $j$, $l = 1, \dots, n_{jk}$ represent the borrower cluster that have the same point estimate, $\hat{Y}_{jk}$, $a_0$ is a global mean parameter, $\alpha_j$ corresponds to school-district-level random effects, and $\alpha_{jk}$ corresponds to school-level random effects, the relationship groups for a point estimate $\hat{Y}_{jk}$ could consist of two clusters: 1) lenders in the same school district but different schools $Y_{jk'}$ (with $a_0$ and $\alpha_j$ in common); and 2) lenders in different school districts $Y_{j'k'}$ (with only $a_0$ in common). The most helpful partition will vary, depending on the model and data.   


To identify which lenders contribute most to SSBF and have the highest individual weight placed on them, it can be helpful to decompose the SSBF into the sum of square borrowing factors over a set of lenders, denoted by $J$, which we call the partial SSBF (PSSBF),  
\begin{equation}
    \label{ssbf-eqn:pssbf}
    \text{PSSBF}_{iJ} = \sum_{j \in J}w_{ij}^2.
\end{equation}
As SSBF is additive, the sum of partial SSBFs over all relationship groups is the SSBF. PSSBF offers a more granular interpretation of SSBF and a scatter plot of partial SSBF against SSBF, colored by relationship group, can identify which {group of} lenders contribute the most to SSBF and thus have the most distinct borrowing patterns, an example of which is in Figure~\ref{ssbf-fig:srd-ssbf}.

Table~\ref{ssbf-tab:defns} repeats and summarizes the definitions for each of the terms listed above. Each is a different way of summarizing information borrowing for a given point estimate $\hat{Y}_i$. When referred to without the subscript $i$, all terms except for $B_i$ and $L_i$ in the table refer to their $N$-length vector counterparts, where the $i^{th}$ entry is, for example, $b_{iJ}$ or $b_{iL_i}$.
\begin{table}[!ht]
    \centering
        \caption{Summary of term definitions and notation for borrowing factors and SSBF of a given point estimate $\hat{Y}_i$, for $i \in \{1, \dots, N\}$. }
\begin{tabular}{lll}
        \toprule
        Notation & Term & Definition \\
        \midrule
        $w_{ij}$ & individual borrowing factor & $(i, j)^{th}$ entry of W \\
        $b_{iJ}$ & aggregate borrowing factor & $\sum_{j \in J} w_{ij}$, for $J \subset \{1, \dots, N\}$ \\
        $B_i$ & borrowers, borrower cluster & $\{j \in 1, \dots, N: x_j = x_i, \phi_{j} = \phi_{i}\}$ \\
        $L_i$ & lenders & $\{1, \dots, N\} \setminus B_i$ \\
        $b_{iB_i}$ & shrinkage factor & $\sum_{j \in B_i}w_{ij}$ \\ 
        $b_{iL_i}$ & pooling factor & $\sum_{j \in L_i}w_{ij}$ \\ 
        $\text{SSBF}_i$ & SSBF & $\sum_{j \in L_i}w_{ij}^2$ \\
        $\text{PSSBF}_{iJ}$ & partial SSBF over $J$ & $ \sum_{j \in J}w_{ij}^2, J \subset L_i$ \\
        \bottomrule
    \end{tabular}
    \label{ssbf-tab:defns}
\end{table}

Exploring the partial SSBF, SSBF, and borrowing factors can help link model assumptions and the data availability to point estimates. 
Depending on the relationship groups, the data, and the model, comparing borrowing factors directly to measures of interest can become quite complex, so we propose a two-stage process. First we determine what contributes the most to changes in SSBF. We compare SSBF to the data availability, model covariates, or some other metric of interest, such as partial SSBF. We then decompose model estimates into borrowing factors over relationship groups and compare them to SSBF, typically as a scatter plot. We can then interpret the change in borrowing factors as SSBF increases or decreases as being due to the data availability, model covariates, or some other metric of interest.We have found this approach to be helpful across different models and data sets and demonstrate it in Sections \ref{ssbf-sec:radon} and \ref{ssbf-sec:srd}. 

\section{Theoretical properties}\label{ssbf-sec:theorems}

We illustrate theoretical properties of the borrowing factors and SSBF. Section~\ref{ssbf-sec:thms_bf} presents properties relevant to the borrowing factors while Section~\ref{ssbf-sec:thms_ssbf} presents properties of SSBF.

\subsection{Properties of the borrowing factors}\label{ssbf-sec:thms_bf}
We first show that the borrowing factors are connected to the shrinkage and pooling factors of the Bayesian literature. Given data $\bm{Y}_i \in \RR^{n_i} \sim N(\alpha_i, \phi_i^2)$, $\alpha_i \sim N(a_0, \sigma^2)$, $i = 1, \dots, J$, where $a_0$, $\phi_i$, and $\sigma$ are known, then it can be shown that the posterior mean, $\hat{Y}_i$, is a balance between the data mean $\bar{Y}_i$ and global mean parameter $a_0$,
\begin{equation}\label{ssbf-eqn:oneway-morris}
    \hat{Y}_i = \lambda_i \bar{Y}_i  + (1 - \lambda_i)a_0, \quad \lambda_i = \frac{\phi_i^2}{n_i\sigma^2 + \phi_i^2},
\end{equation}
where $\lambda_i$ is referred to as the pooling factor and $1 - \lambda_i$ as the shrinkage factor \citep{gelman2006bayesian, efron1975data, morris1983parametric}. This has been used to understand information pooling in Bayesian hierarchical models; clusters with less noise $\phi_i^2$ or more data ($n_i$ is large) borrow less from other points, while those that borrow more are shrunk towards the shared global mean $a_0$. However, this understanding is limited in at least two ways: 1) it is limited to the one-way setting and cannot take into account information borrowing for models with multiple levels and 2) in most cases, $a_0$ is not known and is also informed by $\bar{Y}_i$, so the shrinkage factor in this setting underestimates the total weight placed on $\bar{Y}_i$. As (\ref{ssbf-eqn:oneway-morris}) shows that all point estimates are shrunk towards the global mean $a_0$, it is of interest to understand with more granularity how the data availability or $\phi_i^2$ affects the estimation of $a_0$. 

By conditioning only on the variance parameters, we obtain the borrowing factors, defined in (\ref{ssbf-eqn:w}), and can decompose $\lambda_i$ into weights on each of the data cluster means $\bar{Y}_j$,  
\begin{align}\label{ssbf-eqn:oneway}
\hat{Y}_i &= \left(\frac{n_i\sigma^2 }{n_i\sigma^2 + \phi_i^2} + \rho_{ii}\right)\bar{Y}_i + \sum_{j \neq i}\rho_{ij}\bar{Y}_j,\\
    \rho_{ij} &= \frac{\phi_i^2}{n_i\sigma^2 + \phi_i^2}\frac{\tau_{j}}{\sum_{j = 1}^J\tau_j}, \quad \tau_j \coloneqq \frac{n_j}{n_j\sigma^2 + \phi_j^2}. \notag
\end{align}
It shows that $\lambda_i = \sum_{j = 1}^J \rho_{ij}$. For derivation, see Appendix~\ref{ssbf-appx:onewaybfs}. Instead of one weight placed on the mean parameter $a_0$, which might not be known, we have $J$ borrowing factors which are placed on sample data means $\bar{Y}_j$. This allows us to more closely examine the contribution of $\bar{Y}_j$ to the global mean $a_0$, and thus to $\hat{Y}_i$. This contribution is summarized by $\tau_j$, which is monotonically increasing {in $n_j$} for $n_j \ge 1$ and monotonically decreasing in $\phi_j$. So, the more informative $\bar{Y}_j$ is, with larger $n_j$ or lower noise variance $\phi_j^2$, the closer $\tau_j$ is to its limit, $\sigma^{-2}$. 
{Note $\tau_j \to \sigma^{-2}$ as $\sigma^2$ increases; and, as {$\tau_j$ is finite, as $J$ increases, the input of individual $\tau_j$ lessens in comparison to $\sum_j \tau_j$ and $\tau_j/\sum_j\tau_j \rightarrow 0$.} So, when $\sigma^2$ or $J$ is large, the contribution of any individual $\bar{Y}_j$ to the global mean $a_0$ is low, even under data imbalance.} 

Note that, as defined in (\ref{ssbf-eqn:sf}), the shrinkage factor from (\ref{ssbf-eqn:oneway}) is {the total weight placed on the borrower cluster mean $\bar{Y}_i$ which is} $(1 - \lambda_i) + \rho_{ii}$ and the pooling factor is $\lambda_i - \rho_{ii}$. The $\rho_{ii}$ term accounts for the contribution of $\bar{Y}_i$ to the estimation of global mean parameter $a_0$ and so is moved from $\lambda_i$ to $1-\lambda_i${, where $\lambda_i$ and $1-\lambda_i$ are the shrinkage factor the pooling factor when $a_0$ is known}. 

Having shown that the borrowing factors are equivalent to $\lambda_i$ and $1 - \lambda_i$ in the one-way case, we now show that the properties of the shrinkage and pooling factors in the one-way case generalize to all regression models. In Theorem~\ref{ssbf-thm:sumsto1}, we show that all weights sum to 1. As the borrowing factors can be negative, we additionally show in Theorem~\ref{ssbf-thm:sfunder1} that both the shrinkage and pooling factors are always positive and less than 1 for all regression models. 

\begin{theorem} 
\label{ssbf-thm:sumsto1}
Let response vector $\bm{Y} \in \RR^N$ of a {hierarchical} linear regression follow a normal distribution as in (\ref{ssbf-eqn:normlinreg}), where the $N$-length vector of ones is in the column span of $X_1$, $\mathbf{1} \in \text{span}(X_1)$. In the Bayesian setting, we assume $f(\Sigma)$ and $f(\phi)$ are some prior densities such that the posterior is proper. The $N \times N$ matrix of borrowing factors, $W$, is as defined as in (\ref{ssbf-eqn:w}). Then the sum of borrowing factors $\sum_{j=1}^N w_{ij}$ for a point estimate $\hat{Y}_i$ is 1 for all $i = 1, \dots, N$, i.e. $W\mathbf{1} = \mathbf{1}$.
\end{theorem}

\begin{theorem}\label{ssbf-thm:sfunder1}
Under the same setting as in Theorem~\ref{ssbf-thm:sumsto1}, let the shrinkage factor be defined as in (\ref{ssbf-eqn:sf}). Then given a point estimate $\hat{Y}_i$,  $0 < b_{iB_i} \le 1$ and likewise $0 <= b_{iL_i} < 1$, where $b_{iB_i}$ is the shrinkage factor and $b_{iL_i}$ the pooling factor.
\end{theorem}

For proofs, see \ref{ssbf-appx:sums1prf} and \ref{ssbf-appx:prfsfunder1}, respectively. 
Point estimates $\hat{Y}_i$ can then be seen as balancing the proportion of information coming from the borrower cluster, $b_{iB_i}$, with the proportion of information from the lenders, $b_{iL_i}$.

\subsection{Properties of SSBF}\label{ssbf-sec:thms_ssbf}
We propose to summarize a point estimate's pattern of information borrowing from lenders using the sum squares of borrowing factors (SSBF), defined in (\ref{ssbf-eqn:ssbf}). SSBF has a number of properties that make it suitable for this purpose.
It is lower-bounded by a function of the pooling factor and its relationship to the sample variance of borrowing factors helps to interpret and determine higher-level patterns of information borrowing. SSBF and PSSBF are also related to both model uncertainty and metrics of influence analysis. So, they can be thought of as a more granular version of leverage that summarizes the influence lenders have on a particular $\hat{Y}_i$ due to only the data availability.  In this section, we illustrate and discuss each of these properties.

\vspace{0.5em}

SSBF linearly increases with the sample variance of borrowing factors. Let $b_{iL_i}$ be the pooling factor for $\hat{Y}_i$ and $n_{L_i}$ the number of lenders, then the sample variance is $(n_{L_i} - 1)^{-1}\sum_{j \in L_i}(w_{ij} - {b_{iL_i}}/{n_{L_i}})^2$ and 
\begin{equation}\label{ssbf-eqn:ssbfvar}
\text{SSBF}_i = \sum_{j \in L_i}(w_{ij} - {b_{iL_i}}/{n_{L_i}})^2 + \frac{(n_{L_i}-1)}{n_{L_i} }b_{iL_i}^2.
\end{equation}
For a fixed $b_{iL_i}$, a larger sample variance indicates more distinctive patterns of information borrowing, where some subset of lenders have higher individual borrowing factors than others. SSBF is lowest when a point estimate borrows equally from all lenders. By splitting SSBF into a set of partial SSBFs, as defined in (\ref{ssbf-eqn:pssbf}), we can identify which groups of lenders have consistently high individual borrowing factors. In extreme cases, disproportionately large individual weight may be placed on a few lenders, meaning a large portion of the point estimate is derived from a handful of lenders. As such, researchers may wish to examine such point estimates with high SSBF more closely. 

\vspace{0.5em}

SSBF has a lower bound. Both terms on the right in (\ref{ssbf-eqn:ssbfvar}) are non-negative. The second term then represents a lower bound for the SSBF. Thus SSBF increases as the pooling factor, $b_{iL_i}$, increases. When all borrowing factors are non-negative, as in (\ref{ssbf-eqn:oneway-morris}), SSBF has an upper bound. Using the triangle inequality,
\begin{equation*}
    \sum_{j \in L_i} w_{ij} = b_{iL_i} \implies \sum_{j \in L_i} w_{ij}^2 \le b_{iL_i}^2.
\end{equation*}

\vspace{0.5em}


\vspace{0.5em}

SSBF is related to uncertainty for $\hat{Y}_i$. Let response vector $\bm{Y}$ follow a normal linear regression as in (\ref{ssbf-eqn:normlinreg}) and let $\bm{Y}$ be grouped into clusters $\bm{Y}_i \in \RR^{n_i}$, $i = 1, \dots, J$ such that $\bm{Y}_i \sim N(x_i'\bm{\beta}, \phi_i^2)$. Then the point estimate $\hat{Y}_i$ is a weighted sum over clusters of data means, 
\begin{equation*}
  \hat{Y}_i = \sum_{i = 1}^Jb_{ij}\bar{Y}_j,
  \end{equation*}
as the same weight is placed on all individual points in $\bm{Y}_j$. Let $w_{ij} = b_{ij}/n_j$ denote the individual weight placed on a point in cluster $j$. Knowing only $w_{ij}$ and $\bar{Y}_j$, the central limit theorem states that, for large $n_j$, $\bar{Y}_j \approx N(x_j'\bm{\beta}, \psi_j^2/n_j)$, for some variance $\psi_j^2$. The variance of $\hat{Y}_i$ is then
\begin{equation*}
    \sum_{j}w_{ij}^2n_j\psi_j^2 =  b_{iB_i}^2/n_i\psi_i^2 + \sum_{j\neq i}\text{PSSBF}_{ij}\psi_j^2.
\end{equation*}
Higher SSBF then indicates higher uncertainty surrounding $\hat{y}_i$. This is intuitive when linked to how SSBF is proportional to the sample variance and so larger SSBF values indicate that $\hat{Y}_i$ is borrowing heavily from a relatively small number of lenders. $\hat{Y}_i$ is then more dependent on a smaller set of data points and thus has larger uncertainty. Note, however, that the standard error for $\hat{Y}_i$ also depends on $\psi_j$, thus SSBF is not a direct measurement of uncertainty but summarizes the uncertainty that is due to the data availability.

\vspace{0.5em}

SSBF summarizes the total influence, due to data availability, of all lenders on a point estimate. Influence analysis examines those data points which may have a strong effect on the model fit, without which model parameters could be significantly different. This can be determined through cross-validation, withholding small sets of individual data points at a time. Metrics of influence analysis that are based on single-case-deletion cross-validated estimators have been developed, such as Cook's distance \citep{cook1977detection}. Here we discuss two more recent influence analysis metrics in the literature,  \citeauthor{parsons2018value}'s (\citeyear{parsons2018value}) retrospective value of sample information and \citeauthor{pena2005new}'s (\citeyear{pena2005new}) influence metric $S_i$, and their relationship to PSSBF. 

Value of information is an approach to outlier and influence analysis within the Bayesian literature that quantifies the value of sample information $Y_j$ using the reduction in loss that results from including $Y_j$ v.s. excluding it. Let response vector $\bm{Y}$ follow a normal linear regression as in (\ref{ssbf-eqn:normlinreg}) and let $\bm{Y}_j \in \RR^{n_j} \sim N(x_j'\bm{\beta}, \phi_j^2)$. The retrospective value of sample information (RVSI) of $Y_j$ on $\hat{Y}_i$ can be approximated as the product of the sum of squared residuals and PSSBF,
\begin{equation}\label{ssbf-eqn:rvsi_pssbf}
   \text{RVSI}(Y_j | Y_{-j}; \hat{Y}_i) = \frac{\text{PSSBF}_{ij}}{b_{jL_j}^2}  \frac{n_j(\hat{Y}_j - \bar{Y}_j)^2}{\phi^4}(1 + O(P_2^{-1})).
\end{equation}
For derivation, see Appendix~\ref{ssbf-appx:rvsi}. Partial SSBF is then the portion of the total influence $Y_j$ has on $\hat{Y}_i$ that is due only to the data availability and model definition only, scaled by the squared pooling factor, $b_{jL_j}^2$. 

SSBF has a similar relationship to \citeauthor{pena2005new}'s $S_i$ in the Frequentist literature. \citeauthor{pena2005new}'s $S_i$ is the squared norm of the standardized vector $\bm{s}_i = (\hat{Y}_i - \hat{Y}_{i(1)}, \dots, \hat{Y}_i - \hat{Y}_{i(N)})'$, where $\hat{Y}_{i(j)}=E[Y_i | Y_{-j}]$. $S_i$ has been shown to be able to identify clusters of high-leverage outliers that can be difficult to detect using the usual influence statistics, such as in large high-dimensional data sets. $S_i$ is the sum total of impact all points (lenders and borrowers) have on a point estimate $\hat{Y}_i$. If $\bm{Y}_j \in \RR^{n_j} \sim N(x_j'\bm{\beta}, \phi_j^2)$, $S_i$ can be written as a linear combination of Cook's distances multiplied by the PSSBF,
\begin{equation*}\label{ssbf-eqn:si_pssbf}
    S_i = \sum_{j}\frac{\text{PSSBF}_{ij}}{w_{ii}w_{jj}}\bar{D}_j, \quad \bar{D}_j = \frac{\bar{e}_j^2}{ps^2}\frac{w_{jj}}{(1 - w_{jj})^2},
\end{equation*}
where $D_j$ is the average Cook's distance for $\bm{Y}_j$, $\bm{e} = \bm{Y} - X\bm{\hat{\bm{\beta}}}$, $\bar{e}_j^2 = \bm{e}_j'\bm{e}_j/n_j$,  $s^2 = \bm{e}'\bm{e}/(N -P)$, and $P$ is the dimension of $\bm{\beta}$. See Appendix~\ref{ssbf-appx:si} for derivation.

Note the similarity between RVSI and $S_i$--both can be decomposed into a component describing influence of $Y_j$ due to the data availability and a component describing influence due to squared error $e_j$. For both, PSSBF has a similar role as leverage does to Cook's distance, except it describes the influence of a lender on point estimates. This can be seen by noting that in $S_i$, $\text{PSSBF}_{ij}/w_{ii}$ replaces the leverage term $w_{jj}$ that is in $\bar{D}_j$ and in RVSI, if $n_j = 1$, then $b_{jL_j}^2 = (1 - w_{jj})^2$. 


These properties make SSBF and PSSBF helpful metrics for summarizing how a point estimate $\hat{Y}_i$ borrows from its lenders. Higher SSBF indicates the point estimate may borrow more from a small number of lenders and therefore has more distinct borrowing patterns. Examining those points with high SSBF can help researchers identify borrowing patterns that are crucial for model estimates. 



\section{Example: Radon}\label{ssbf-sec:radon}

We demonstrate how SSBF and the borrowing factors can explain the impact of data imbalance on model estimates and information borrowing. The Radon data measures the log radon level of 919 houses in Minnesota and contains data on the house's county, the average level of uranium in the county, and whether the house contains a basement. The data are included as part of the \texttt{rstanarm} package \citep{gabry2016rstanarm} via \citet{gelman2007data}. 

We model the log radon level of houses in county $j$ and basement status $k$  with a fixed effect intercept $a_{0k}$ based on basement status, fixed effect coefficient ${a_1}$ using the log uranium value, and county-specific random intercept $\alpha_j$:
\begin{align}
    \bm{Y}_{kj} \sim &N({a}_{0k} + {a_1} u_j + \alpha_j, \phi^2) \label{ssbf-eqn:radonsetup}\\
    &{a}_{0k} \sim N(0, c_k^2), {a_1} \sim N(0, c^2) \notag \\
    &\alpha_j \sim N(0, \sigma^2), \sigma \sim f(\sigma), \phi \sim f(\phi), \notag
\end{align}
where $u_j$ is the \texttt{log uranium} value for county $j$, $c_k$ and $c$ are fixed scalar values $\in \RR^+$, representing the variances of $a_{0k}$ and ${a_1}$ respectively, and $\alpha_j$ denotes county-specific random effects. The model was fit using \texttt{rstanarm}, using the default priors and hyperparameters for \texttt{stan\_lmer}, under which $c_k = 2$, $c = 5.5$, $\phi \sim \textup{Exp}(1)$, and $\sigma \sim \textup{Exp}(1)$.

The data are imbalanced across counties and basement status. There are 85 total counties with a mean of 10.8 houses per county, a median of 5, and inter-quartile range from $3$ to $10$. The eight counties with the most houses make up 50\% of the data set. Two of the counties contain data on over 100 houses, each making up over 11\% of the data. 766 of the houses (83\%) do not have a basement and 153 (17\%) do. Intuitively, one would expect that counties with fewer houses borrow more from the counties with a larger number of houses. 
The borrowing factors allow us to explicitly quantify the amount of borrowing for each county and link this to the data availability. 
For this example, we partition the observations into the following relationship groups:
\begin{itemize}
    \item the borrower cluster $\bm{Y}_{kj} \in \RR^{n_{kj}}$,
    \item same-county lenders $\bm{Y}_{k'j} \in \RR^{n_{k'j}}$, 
    \item same-basement lenders $\bm{Y}_{kj'} \in \RR^{n_{kj'}}$,
    \item lenders in a different county with a different basement status $\bm{Y}_{k'j'} \in \RR^{n_{k'j'}}$,
\end{itemize}

\begin{figure}[!ht]
    \centering
    \includegraphics[height=10em]{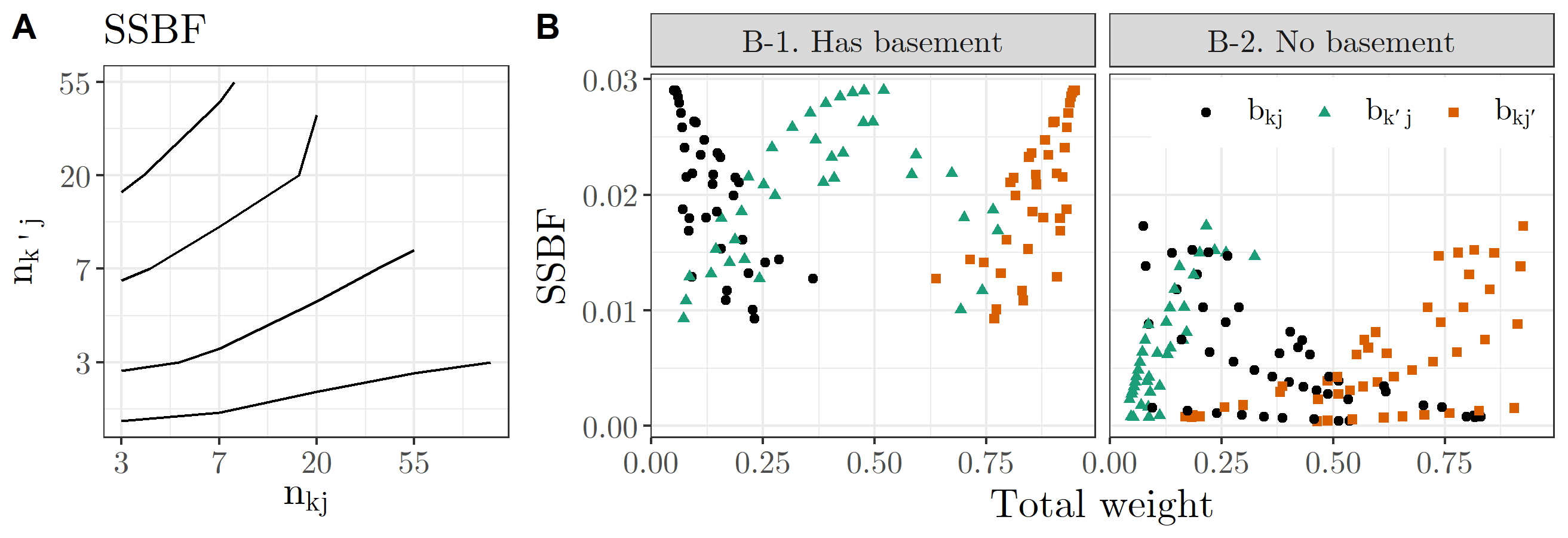}
    \caption{For the Radon data, modeled as in (\ref{ssbf-eqn:radonsetup}). Panel A is a contour plot of SSBF; 
    contours are based on the mean SSBF for each unique combination of borrower cluster size $n_{kj}$ and same-county lender size $n_{k'j}$. Panel B is a scatter plot of SSBF against the shrinkage factor ($b_{kj}$) and two borrowing factors corresponding to lenders in the same county and different basement status ($b_{k'j}$) and lenders with the same basement status ($b_{kj'}$).}
    \label{ssbf-fig:radon}
\end{figure}

We first compare SSBF to measures of data availability. Figure \ref{ssbf-fig:radon}A is a contour plot of SSBF with the borrower cluster size $n_{kj}$ and the number of same-county lenders $n_{k'j}$ on the x- and y-axes.  As $n_{k'j}$ increases, SSBF increases, which implies that lenders in the same county have large individual weights placed on them. As $n_{kj}$ decreases, SSBF increases, showing that more is borrowed from same-county lenders to compensate for low borrower cluster size. When $n_{k'j} = 0$, SSBF is low regardless of $n_{kj}$, indicating that none of the remaining lenders has particularly high individual weight placed on them. Borrowing within the same county is then the most distinctive pattern of borrowing that changes with the data availability and is the main contributor to the change in SSBF across data points. 

Next, we examine the borrowing factors for the three relationship groups defined earlier. As we are mainly interested in the effects of data availability which corresponds to the basement status and the county effects, we consider the point estimates conditional on $a_1$. Let $\hat{\mu}_{kj} \coloneqq E[a_{0k} + \alpha_j | a_1, \bm{Y}] = \hat{Y}_{kj} - \hat{{a}}_1 u_j$, $b_{kj}$ be the shrinkage factor for $\hat{\mu}_{kj}$, $b_{k'j}$ be the total amount borrowed from $\bm{Y}_{k'j}$, and $b_{kj'}$ be the total amount borrowed from $\bm{Y}_{kj'}$. 

We notice that $b_{k'j} = -b_{k'j'}$ and only present the borrower cluster and the first two relationship groups. Appendix~\ref{ssbf-appx:bal} provides intuition for why $b_{k'j} = -b_{k'j'}$.
Figure \ref{ssbf-fig:radon}B compares the shrinkage factor $b_{kj}$ and borrowing factors $b_{k'j}$ and $b_{kj'}$ to SSBF for all point estimates $\bm{\hat{\mu}}$. Note that $b_{kj}$ and $b_{kj'}$ are reflections of each other across a vertical line at $0.5$ and thus $(b_{kj} + b_{kj'}) = 1$ for all data points. This is because $(b_{kj} + b_{kj'} + b_{k'j} + b_{k'j'}) = 1$ and the summation of the last two terms is zero, as noted earlier. As $n_{kj}$ increases, $b_{kj} \rightarrow 1$ and $b_{kj'} \rightarrow 0$, and vice versa. Borrowing via the county intercept occurs through relationship groups with the same $j$ and is represented by $b_{kj} + b_{k'j}$. This quantity is typically less than 1. When the number of houses in the county, $n_j$, is large, $b_{kj} + b_{k'j}  \rightarrow 1$ and when $n_j$ is small, $b_{kj} + b_{k'j}$ is small, i.e., the model will shrink the amount of borrowing via the county intercept. {Thus $b_{kj} + b_{k'j}$ quantifies the impact of $n_j$ on model estimates. This can be further decomposed into the impacts of $n_{kj}$ and $n_{k'j}$, using borrowing factors $b_{kj}$ and $b_{k'j}$.}

In Figure \ref{ssbf-fig:radon}A, we saw that $n_{k'j}$ is closely related to the SSBF and typically increases as SSBF increases. In Figure \ref{ssbf-fig:radon}B, that relationship in more detail.  As $b_{k'j}$ increases towards 0.5, $n_{k'j}$ increases and so does SSBF. For higher values of $n_{k'j}$ and $b_{k'j}$, SSBF begins to decrease again as the larger number of data points means no single data point gets a large weight.

One model assumption is that all houses in a county are equally informative of the county-specific effect. As such, the borrowing factors weight both $Y_{kj}$ and $Y_{k'j}$ nearly equally---for the point in panel A with highest SSBF, $b_{k'j} = 0.5$ and $b_{kj} = 0.05$, while $n_{k'j} = 12$ and $n_{kj} = 1$. (The slight difference is because $Y_{kj}$ is also informative for the floor effect, but as there are many other points to inform the floor effect, it is not necessary to place high additional weight on $Y_{kj}$.) In other words, $Y_{k'j}$ has much higher total weight placed on it than the borrower cluster's own data, $Y_{kj}$. This is the case for many of the points in panel B-1, where the shrinkage factor is typically under 0.25 but most $b_{k'j}$s are over 0.25. This is due to the data availability, where fewer houses have basements and so more information is borrowed from those that do. It follows that the reverse is the case in panel B-2, where $n_{kj}$ is typically larger than $n_{k'j}$ and, as such, many of the point estimates have shrinkage factor over 0.25 with most $b_{k'j}$s are under 0.25. Overall, the point estimates with low shrinkage factor and high $b_{k'j}$ are the most affected by this model assumption and are also the counties with the highest data imbalance across basement status. 

By comparing SSBF to the data availability in Figure~\ref{ssbf-fig:radon}A, we determined that the number of lenders in the same county is the main contributor to the change in borrowing patterns across data points. By comparing SSBF to the borrowing factors in Figure~\ref{ssbf-fig:radon}B, we were able to link the data availability and model assumptions to patterns of information borrowing. Much of this was intuitive. The borrowing factors simply allow us to place explicit numbers on the degree to which point estimates are affected. 
In scenarios with more complex models or more severe data imbalance, the intuition may not be so readily available, but the borrowing factors and SSBF can still tell us which point estimates borrow the most from others and which points they borrow from.


\section{Example: Scottish respiratory disease}\label{ssbf-sec:srd}

Here, we examine a more complex Bayesian hierarchical generalized linear model with spatio-temporal conditional auto-regressive (CAR) intercepts. In Section~\ref{ssbf-sec:srd_mod}, we identify the data properties which contribute to higher SSBF and high-level patterns of information borrowing. In Section~\ref{ssbf-sec:srd_infl}, we demonstrate how this understanding of model estimates can be used to provide context to influence analysis. 

The Scottish respiratory disease data consists of annual observed respiratory-related hospital admissions in the $J = 271$ Intermediate Geographies (IG) of the Greater Glasgow and Clyde health board from 2007 - 2011; the yearly average modelled concentrations of particulate matter less than 10 microns (\texttt{$\text{PM}_{10}$}); the average property price in hundreds of thousands of pounds (\texttt{Property}); the proportion of the working age population who receive an unemployment benefit called the Job Seekers Allowance (\texttt{JSA}); the expected number of hospital admissions, $E_{tj}$, which is modeled as an offset-term; and the adjacency matrix $A$, where $A_{ii} = 0$, $A_{ij} =  A_{ji} = 1$ if $j$ and $i$ are neighboring districts, and $0$ otherwise. It is available through the \texttt{CARBayesST} package in R. 

We use the spatio-temporal auto-regressive model in \citet{rushworth2014spatio}, where observed hospital admissions for a year $t$ and IG $j$ are modelled with a Poisson density,
\begin{align*}
    Y_{tj} &= \text{Poisson}(\eta_{tj}E_{tj}) \\
    \log(\eta_{tj}) &= x_{tj}'\bm{a} + \alpha_{tj},
\end{align*}
where $x_{tj}$ is a vector containing \texttt{$\text{PM}_{10}$}, \texttt{Property}, and \texttt{JSA} values for that year $t$ and IG $j$; and $\bm{a}$ is the vector of fixed effects. Within each year, spatial dependence among the corresponding vector of random effects $\bm{\alpha}_t = (\alpha_{t1}, \dots, \alpha_{tJ})'$ is modeled with covariance matrix $\sigma^2Q(\rho_J, A)^{-1}$, where 
\begin{equation*}
    Q(\rho_J, A)^{-1} = \rho_J(\text{diag}(W\mathbf{1}) - A) + (1 - \rho_J)I_J, \quad \rho_J \in [0, 1),
\end{equation*}
which induces spatial auto-correlation and is a special case of a CAR model. Temporal auto-correlation is introduced among the $\alpha_t$ by the conditional density of $\alpha_t | \alpha_{t-1}$:
\begin{equation*}
    \alpha_t | \alpha_{t-1} \sim N(\rho_T \alpha_{t-1}, \sigma^2 Q(\rho_J, A)^{-1}), j \in \{2, \dots, T\}.
\end{equation*} 
The model is fit using the \texttt{ST.CARar()} function in \texttt{CARBayesST} with the default priors $\bm{a} \sim N(0, 100,000)$, $\sigma \sim IG(1, 0.001)$, $\rho_T \sim U(0, 1)$, $\rho_J \sim U(0, 1)$. The resulting posterior means for spatial dependence parameter $\rho_J$ and temporal dependence parameter $\rho_T$ are $0.57$ and $0.76$, respectively.

As the data are modeled with a Poisson GLMM, the normal priors are not conjugate and the analytical form of (\ref{ssbf-eqn:axe_linregr}) is no longer available. We instead approximate the data-level Poisson model with a normal distribution having equivalent moments, as described in \citet{daniels1998note}, maintaining conjugacy and a closed-form solution for the borrowing factors. Sample sizes within this data set were large enough that the normal approximation produced closely similar estimates when we compared the normal approximation to actual posterior means (see Appendix~\ref{ssbf-appx:srd-figs}). In this case, our approximating normal density is 
\begin{equation}\label{ssbf-eqn:srd-normapprox}
      \log(Y_{tj}) - \log(E_{tj}) | \eta_{tj}, E_{tj} \approx N\left(\log(\eta_{tj}), \eta_{tj}^{-1}\right)
\end{equation}

and we can obtain SSBF along with borrowing factors as described in (\ref{ssbf-eqn:ssbf}). We derive the joint density of $\bm{\alpha} = (\alpha_1', \dots, \alpha_T')'$:
\begin{align*}
  s &\sim N(0, \sigma^{2}[(I - \rho_T H)\text{blockdiag}(Q(\alpha, W))(I - \rho_T H)]^{-1}) \\
    H &= \begin{bmatrix}
    \bm{0}_{J \times J(T - 1)} & \bm{0}_{J \times J} \\
   {\bm{I}_{J(T - 1)}} & \bm{0}_{J(T - 1) \times J} 
    \end{bmatrix},
\end{align*}
where $I_{J(T-1)} \in \RR^{J(T-1) \times J(T-1)}$ is the identity matrix, and $\bm{0}$ are matrices of 0s with dimensions such that $H\in \RR^{JT\times JT}$ accounts for the temporal auto-correlation.

For this model, we aggregate the borrowing factors and partial SSBF based on how close the lender is to the borrower, which can be defined both temporally and spatially. The relationship groups are combinations of three spatial and three temporal categories, where the spatial categories are
\begin{itemize}
    \item the lender is in the same IG, denoted with subscript $j_0$,
    \item the lender is in a neighboring IG ($j_1$),
    \item or the lender is farther away ($j_{2+}$),
\end{itemize}
and the temporal categories are
\begin{itemize}
    \item the lender is in the same year, denoted with subscript  $t_0$,
    \item the lender is in 1 year away ($t_1$),
    \item the lender is 2 or more years away ($t_{2+}$),
\end{itemize}
resulting in 9 total relationship groups. 

\subsection{High-level information borrowing patterns}\label{ssbf-sec:srd_mod}

From the posterior means for spatial dependence parameter $\rho_J$ and temporal dependence parameter $\rho_T$ {($\hat{\rho}_T>\hat{\rho_J}$)}, we may have some intuition that for point estimate $\hat{Y}_{tj}$, $Y_{t_1,j}$  may have higher weight than $Y_{t,j_1}$, but it is not clear how other lender groups affect $\hat{Y}_{tj}$ and whether, for example, $Y_{t_{1},j_1}$ has noticeable impact on $\hat{Y}_{tj}$ or not. In this section we quantify and compare borrowing across each of the relationship groups to understand which lenders have the most impact on point estimates.

\begin{figure}[!ht]
    \centering
    \includegraphics[width=\linewidth]{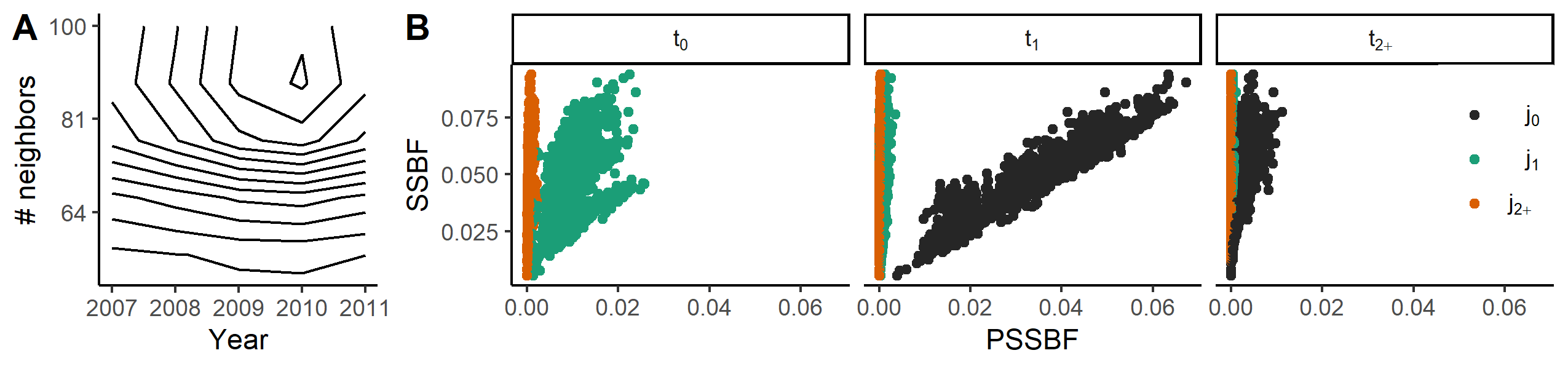}
    \caption{Panel A is a contour plot of smoothed SSBF values against the year and number of neighbors for each point. Smoothing is conducted with a Nadaraya-Watson type kernel estimator. Panel B is a scatter plot of SSBF against partial SSBF, where each panel represents a different temporal relationship group ($t_0$, $t_1$, $t_{2+}$ for same year, adjacent year, other years, respectively) and colors represent different spatial relationship groups (black for $j_0$, green for $j_1$, orange for $j_{2+}$, corresponding to same IG, neighboring IG, and farther IGs, respectively).}
    \label{ssbf-fig:srd-ssbf}
\end{figure}

First, we identify what has the largest impact on SSBF and the borrowing patterns. Figure~\ref{ssbf-fig:srd-ssbf} illustrates this in two ways. The first, in panel A, is a contour plot of SSBF against two properties of the data, the number of neighbors and the year 
(this is similar to the contour plot in Section~\ref{ssbf-sec:radon}, Figure~\ref{ssbf-fig:radon}, which links data availability to the SSBF). The second, in panel B, is a scatter plot of SSBF vs PSSBF which helps to identify which borrowing factors contribute the most to the change in SSBF. 

The contour plot links data properties to SSBF and shows that SSBF is the highest for those points at year 2010 with around 90 neighbors. Those points have more potential lenders to borrow from, with a large number of neighboring IGs and two neighboring time points. The scatter plot is a high-level summary of the borrowing patterns and identifies which borrowing factors change the most with SSBF. If PSSBF has a large positive correlation with SSBF, then it is likely that the lenders in that relationship group have high individual weight placed on them. We can see that the borrowing factors for $\bm{Y}_{t_1j_0}$ (Figure~\ref{ssbf-fig:srd-ssbf}B center panel, black points), $\bm{Y}_{t_0j_1}$ (Figure~\ref{ssbf-fig:srd-ssbf}B left panel, green points), and $\bm{Y}_{t_{2+}j}$ (Figure~\ref{ssbf-fig:srd-ssbf}B right panel, black points) contribute the most to the change in SSBF, in decreasing order of impact. Correlations between PSSBF and SSBF are $0.94, 0.47$ and $0.36$ respectively for each of the relationship groups.

By comparing SSBF to data properties in Figure~\ref{ssbf-fig:srd-ssbf}A, we determined that point estimates with the highest SSBF values were typically those with a large number of neighbors near the year $2010$. The model induces positive correlations on points in neighboring IGs or neighboring years, thus those points that have more neighbors to borrow from have more distinct information borrowing patterns and higher SSBF. We identified which lenders contribute the most to the change in SSBF and thus likely have the highest individual weights placed on them using  Figure~\ref{ssbf-fig:srd-ssbf}B. 
These relationships may not be readily apparent when examining the posterior mean estimates and the data alone, but can be determined by examining the borrowing factors which quantify the relative amounts of information borrowing for each of the relationship groups.

More detailed investigation of the relative magnitude of the borrowing factors for each relationship group can be determined by comparing SSBF to the borrowing factors, as in the \texttt{ssbf} package Shiny app. A plot of SSBF against borrowing factors is included in the supplementary material, Appendix~\ref{ssbf-appx:srd-figs}.

\subsection{Impact of influential points}\label{ssbf-sec:srd_infl}

Influence analysis examines those data points which may have a strong effect on the model fit, without which model parameters could be significantly different. After identifying influential points through the use of a metric such as Cook's distance, RVSI, or $S_i$, a decision is often made on whether they are outlying, typically based on subject matter considerations and their degree of influence. 
By examining which point estimates rely the most on these influential points, we can add more context to subject matter considerations of whether to keep or discard the influential points and contextualize their degree of influence on other point estimates. Using SSBF and the borrowing factors, we can understand exactly how an influential point $Y_i$ affects other model estimates $\hat{\mu}_j$ and thus identify those estimates that are most impacted by $Y_i$.

We identified a set of 11  potentially influential points $s$ using PCA-decomposition of the log case-deletion importance sampling weights, as described in \citet{thomas2018reconciling}, which captures both global case influence of an individual point, in terms of distance from the full-data and the case-deleted posterior, and local case influence, through perturbations to the likelihood. Any method which produces estimates of influence for all data points $\bm{Y}$ can be used. 

A point may be influential because of the data availability; in these cases, the covariates corresponding to the point are unique in some way, such as belonging to a rare category or having extreme values. This is most commonly summarized via leverage, essentially the square root of diagonal values of $W$, where higher values indicate the point has higher impact on model estimates. The point may also be influential because the response value is unexpected in some way under the model. In either case, the points that are most impacted by an influential point are those for which the borrowing factor is higher. 

\begin{figure}
    \centering
    \includegraphics[height=12em]{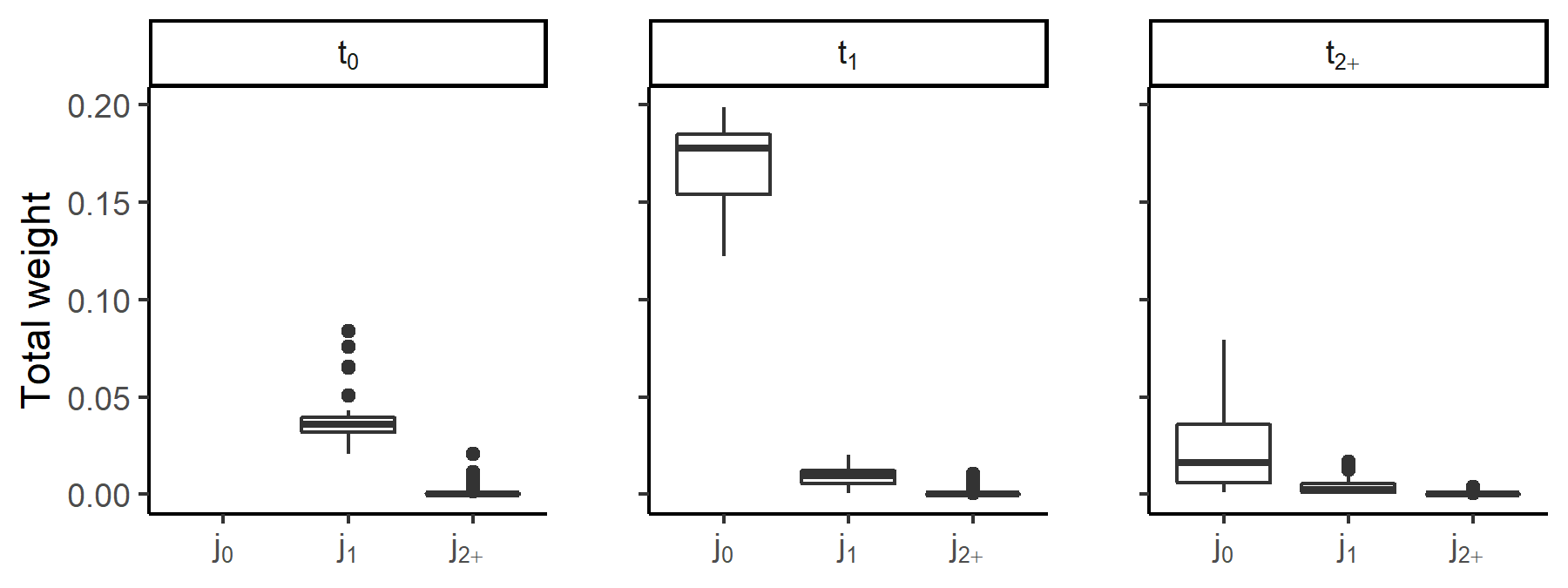}
    \caption{Boxplots of total (absolute) weight placed on 11 influential points, split into temporal ($t_0$, $t_1$, $t_{2+}$) and spatial ($j_0$, $j_1$, $j_{2+}$) relationship groups. The plots do not include the shrinkage factor, hence no boxplot for $b_{t_0j_0}$.}
    \label{ssbf-fig:srd-infl}
\end{figure}

Figure \ref{ssbf-fig:srd-infl} consists of boxplots of individual borrowing factors on the 11 influential points, for all model estimates. The boxplots show that the influential points have the most impact on neighboring time points that are in the same IG, with median borrowing factor near 0.18. {The influential points also have a noticeable impact on point estimates for neighboring IGs in the same year and those in the same IG, but more than 1 year away.} Both typically have borrowing factors under 0.05. Other relationship groups are less affected, with borrowing factors generally near 0. This is in line with the SSBF vs PSSBF plot in Figure~\ref{ssbf-fig:srd-ssbf}B, which shows that individual borrowing factors are low for neighboring IGs at the same year. Part of this could be because the temporal dependence is larger than the spatial dependence, based on posterior samples, but a large part of this is likely due simply to data availability. Plots of SSBF against the borrowing factors show that borrowing factors for neighboring IGs at the same year and neighboring years at the same IG are similar in magnitude (see Appendix~\ref{ssbf-appx:srd-figs}). There are typically a large number of neighboring IGs to borrow from, so less individual weight is placed on each neighbor, lessening the impact of any individual point. There are only one or two neighboring time points that are at the same IG, which leads to higher individual weight placed on those timepoints. We can conclude that although both spatial and temporal dependence in the model is high, influential points will have much greater impact on point estimates from neighboring time points because of the data availability. This can be confirmed by obtaining the weights if the posterior means for $\rho_T$ and $\rho_J$ are switched so that $\rho_J = 0.76$ and $\rho_T = 0.57$, which results in a similar boxplot (see Appendix~\ref{ssbf-appx:srd-figs}).  

By decomposing model estimates using the borrowing factors, we explicitly quantify which point estimates are the most and least impacted by by the 11 identified influential points. We determined that those are the point estimates that are next to an influential point in time, with median borrowing factor around 0.18, followed by those point estimates that are in neighboring IGs, with median borrowing factor under 0.05. Based on the conclusions from Section~\ref{ssbf-sec:srd_mod}, we determined that the relatively low borrowing factors on neighboring IGs was due to the data availability.

\section{Discussion}
\label{ssbf-sec:discussion}
Borrowing factors explicitly quantify how the data availability and model specification impact model estimates. We demonstrated this with two examples. In the Radon example, we used both borrowing factors and SSBF over same-county lenders to quantify the impact of data availability on model estimates. In the SRD example, we showed how the number of neighboring lenders affected point estimates and used this understanding to identify lenders that are most impacted by influential points. In both cases, the borrowing factors allowed us to place explicit quantities on relationships that could previously be assumed but would be difficult to verify.

We examined the properties of borrowing factors for point estimates, $\hat{Y}_i$. Researchers may also use the borrowing factors to examine particular coefficients. In this case, the weight  matrix $W$  would then be taken as $VX'\Phi^{-1}Y$.   

As the dimension of $W$ is often large, we encourage graphical summaries to understand the borrowing factors and SSBF. Graphs can be used to identify both high-level patterns among point estimates as well as providing granular information on a single point estimate. We have found that we can understand model estimates by comparing SSBF to the borrowing factors, partial SSBF, measures of data availability, and model covariates.  We provide an R package for creating these plots and an interactive Shiny app for simultaneously displaying multiple plots. Users can select points in any plot, which will then be highlighted and annotated with information across all plots.

With its focus on examining the mechanisms of regression models, philosophically, our approach resembles methods in the explainable machine learning literature,  particularly those which allow for integrating domain knowledge \citep{yan2019groupinn, tsang2018detecting};  see \citet{roscher2020explainable} for a survey and taxonomy of explainable machine learning. The borrowing factors themselves bear the most resemblance in the literature to the pooling factor which, to our knowledge, is the only method in the literature which derives an explicit quantity that describes and quantifies information borrowing.  



\section*{Data Availability Statement}

The datasets and the code for implementing the analysis in this manuscript are available at \texttt{https://github.com/amytildazhang/ssbf}.

\section*{Acknowledgement}
This work was supported by the National Institutes of Health (NIH) under grants R56AI120812-01A1 and R01AI136664.

Supplementary material available at xxx online includes technical details and proofs for the borrowing factors and SSBF, as well as supplementary figures for the Radon and Scottish respiratory disease data examples.

\section*{Supplementary material}
\label{ssbf-SM}
Appendix 1 contains all proofs corresponding to Section~\ref{ssbf-sec:thms_bf} in the paper. Appendix 2 illustrates the relationship between SSBF and influence analysis metrics RVSI and $S_i$, discussed in Section~\ref{ssbf-sec:thms_ssbf}.  Appendix 3 provides further explanation and intuition on why it is sufficient to examine only two borrowing factors in the Radon data example in Section~\ref{ssbf-sec:radon}.

\vspace*{-10pt}

\appendix

\section*{Appendix 1}
\ref{ssbf-appx:onewaybfs} contains the derivation for the borrowing factors under a one-way model. \ref{ssbf-appx:sums1prf} contains the proof for Theorem~\ref{ssbf-thm:sumsto1}. \ref{ssbf-appx:prfsfunder1} contains the proof for Theorem~\ref{ssbf-thm:sfunder1}.

\subsection{Borrowing factors for one-way models}\label{ssbf-appx:onewaybfs}
Here we provide the calculations for the borrowing factors in the one-way setting, shown in (\ref{ssbf-eqn:oneway}).  Given data $\bm{Y}_i \in \RR^{n_i} \sim N(\alpha_i, \phi_i^2)$, $\alpha_i \sim N(\mu, \sigma^2)$, where $\mu \in \RR$, $\alpha_i \in \RR$, and $i = 1, \dots, J$. In this scenario, it is possible to analytically solve for the borrowing factors in (\ref{ssbf-eqn:w}). 

We begin by solving for $V$. Defining $X_1$ and $X_2$ as in (\ref{ssbf-eqn:axe_linregr}), we can write $V^{-1}$ as a block matrix
\begin{equation}\label{ssbf-eqn:vblock}
V^{-1} = 
\left[\begin{array}{@{}cc@{}}
 X_1'\Phi^{-1}X_1 & X_1'\Phi^{-1}X_2 \\ 
 X_2'\Phi^{-1}X_1 & X_2'\Phi^{-1}X_2 + \Sigma^{-1}
\end{array}\right]
\end{equation}
and obtain a solution for $V$ using the rules for block matrix inversion. Starting in the upper-left quadrant and moving clockwise, let us refer to the corresponding blocks of $V$ as $A, B, C, D$, such that
\begin{equation*}
    V = \begin{bmatrix}A & B \\ C & D\end{bmatrix},
\end{equation*}
and $A \in \RR^{P_1 \times P_1}, B \in \RR^{P_1 \times P_2}, C = B' \in \RR^{P_2 \times P_1}, D \in \RR^{P_2 \times P_2}$.  
 
In this scenario, $X_1 = \bm{1}_N$, the vector of ones, and $X_2$ is the binary matrix of indicator variables where the $i^{th}$ column indicates membership in the $i^{th}$ cluster. Then the form of each block is as follows, 
\begin{equation*}
    V^{-1}  = 
    \left[\begin{array}{@{}c|cccc@{}}
        \sum_{i=1}^P \frac{n_i}{\phi_i^2} & \frac{n_1}{\phi_1^2} & \frac{n_2}{\phi_2^2} & \dots & \frac{n_J}{\phi_J^2} \\
        \hline
        \frac{n_1}{\phi_1^2} & \frac{n_1}{\phi_1^2} + \sigma^{-2} &  0 & \dots &   0 \\
        \frac{n_2}{\phi_2^2} & 0 & \frac{n_2}{\phi_2^2} + \sigma^{-2} &  \dots &   0 \\
        \vdots & \vdots & \ddots &  0 \\
        \frac{n_J}{\phi_J^2} & 0 & \dots & 0 & \frac{n_J}{\phi_J^2} + \sigma^{-2}
    \end{array}\right]
\end{equation*},
where the vertical and horizontal lines enclose each of the four blocks in (\ref{ssbf-eqn:vblock}). 

We can now solve for $A$, using the rules for block matrix inversion,
\begin{align*}
    A   &= \left(\sum_{j=1}^J \frac{n_j}{\phi_j^2} - \sum_{j=1}^J \frac{n_j^2/\phi_j^4}{n_j/\phi_j^2 + \sigma^{-2}} \right)^{-1} \\
          &= \left(\sum_j\left(\frac{n_j}{\phi_j^2}\left(1 - \frac{n_j\phi_j^{-2}}{n_j\phi_j^{-2} + \sigma^{-2}}\right)\right)\right)^{-1} \\
          &= \left(\sum_j\left(\frac{n_j}{\phi_j^2}\frac{\sigma^{-2}}{n_j\phi_j^{-2} + \sigma^{-2}}\right)\right)^{-1} \\
          &= \left(\sum_j\left(\frac{n_j}{n_j\sigma^{2} + \phi^2}\right)\right)^{-1} \\
          &= \left(\sum_j \tau_j \right)^{-1},
\end{align*}{}
where $\tau_j \coloneqq n_j/(n_j\sigma^2 + \phi_j^2)$ as in (\ref{ssbf-eqn:oneway}).

We derive the remaining block matrices of $V$ in terms of $\tau_j$ and $A$.
\begin{align*}
   B &= \left\{-\tau_j\sigma^2A \right\}_{1 \times J}, \\
      D &= \text{diag}\left(\frac{\phi_j^2\sigma^2}{n_j\sigma^2 + \phi_j^2}\right) + \left\{\tau_j\sigma^2A\tau_{j'}\sigma^2 \right\}_{J \times J} \\
      &=  \text{diag}\left(\frac{\phi_j^2\sigma^2}{n_j\sigma^2 + \phi_j^2}\right) + \left\{B_jA^{-1}B_{j'}\sigma^2 \right\}_{J \times J} 
\end{align*}{}

With $V$ known, with some algebra, we can derive the final result,
\begin{align*}
 \hat{y_i} &= x_i'VX'\Phi^{-1}\bm{Y} \\
                 &= (A + B_i)\sum_{j}\frac{n_{j}}{\phi_{j}^2}\bar{Y}_{j} + \sum_{j}\frac{n_{j}}{\phi_{j}^2}\bar{Y}_{j}{Bj}\left(1 +  B_iA^{-1}\right) + \tau_i\sigma^2\bar{Y}_i \\
      &= A\frac{\phi_i^2}{n_i\sigma^2 + \phi_i^2}\sum_{j}\frac{n_{j}}{\phi_{j}^2}\bar{Y}_{j} +\sum_{j}\frac{n_{j}}{\phi_{j}^2}\bar{Y}_{j}{Bj}\frac{\phi_i^2}{n_i\sigma^2 + \phi_i^2} + \tau_i\sigma^2\bar{Y}_i \\
      &= A\frac{\phi_i^2}{n_i\sigma^2 + \phi_i^2}\sum_{j}\frac{n_{j}}{\phi_{j}^2}\bar{Y}_{j}\frac{\phi_{j}^2}{n_{j}\sigma^2 + \phi_{j}^2}+ \tau_i\sigma^2\bar{Y}_i \\
      &= \frac{\phi_i^2}{n_i\sigma^2 + \phi_i^2}\sum_{j}\frac{\tau_{j}}{\sum_j\tau_j}\bar{Y}_{j} + \tau_i\sigma^2\bar{Y}_i.
\end{align*}

\subsection{Proof of Theorem~\ref{ssbf-thm:sumsto1}}\label{ssbf-appx:sums1prf}
We re-state Theorem~\ref{ssbf-thm:sumsto1} below for reference:

Theorem 1: Let response vector $\bm{Y} \in \RR^N$ of a {hierarchical} linear regression follow a normal distribution as in (\ref{ssbf-eqn:normlinreg}), where the $N$-length vector of ones is in the column span of $X_1$, $\mathbf{1} \in \text{span}(X_1)$. In the Bayesian setting, we assume $f(\Sigma)$ and $f(\phi)$ are some prior densities such that the posterior is proper. The $N \times N$ matrix of borrowing factors, $W$, is as defined as in (\ref{ssbf-eqn:w}). Then the sum of borrowing factors $\sum_{j=1}^N w_{ij}$ for a point estimate $\hat{Y}_i$ is 1 for all $i = 1, \dots, N$, i.e. $W\mathbf{1} = \mathbf{1}$.

\begin{proof}

Defining $X_1$ and $X_2$ as in (\ref{ssbf-eqn:axe_linregr}), we can write $V^{-1}$ as a block matrix
\begin{equation*}
V^{-1} = \begin{bmatrix}X_1'\Phi^{-1}X_1 & X_1'\Phi^{-1}X_2 \\ X_2'\Phi^{-1}X_1 & X_2'\Phi^{-1}X_2 + \Sigma^{-1}\end{bmatrix}
\end{equation*}
and obtain a solution for $V$ using the rules for block matrix inversion. Starting in the upper-left quadrant and moving clockwise, let us refer to the corresponding blocks of $V$ as $A, B, C, D$, such that
\begin{equation*}
    V = \begin{bmatrix}A & B \\ C & D\end{bmatrix},
\end{equation*}
and $A \in \RR^{P_1 \times P_1}, B \in \RR^{P_1 \times P_2}, C \in \RR^{P_2 \times P_1}, D \in \RR^{P_2 \times P_2}$.

Let $M \coloneqq (X_2'\Phi^{-1}X_2 + \Sigma^{-1})^{-1}$; $H_2 \coloneqq X_2MX_2'\Phi^{-1}$; and $\tilde{\Phi}^{-1} \coloneqq \Phi^{-1}(I - H_2)$. We solve for each of the blocks in $V$ and write the solutions in terms of $M$, $H_2$, and $\tilde{\Phi}^{-1}$:
\begin{align}\label{ssbf-eqn:prfvblocj}
A &= (X_1'\Phi^{-1}X_1 - X_1'\Phi^{-1}H_2'\Phi^{-1}X_1)^{-1} =  \left(X_1'\tilde{\Phi}^{-1}X_1\right)^{-1} \\
B &= -AX_1'\Phi^{-1}X_2M   \notag\\
C &= B'\notag\\
D &= M + MX_2'\Phi^{-1}X_1AX_1'\Phi^{-1}X_2M. \notag
\end{align}

Let $H \coloneqq X_1(X_1'\tilde{\Phi}^{-1}X_1)^{-1}X_1'\tilde{\Phi}^{-1}$ and $H_1 \coloneqq X_1AX_1'{T}^{-1}$. Note that $H = H_1(I - H_2)$. The weight matrix can be re-written in terms of $H$ and $H_2$ using (\ref{ssbf-eqn:prfvblocj}),
\begin{align*}
W &= XVX'^{-1}\Phi^{-1} = X_1AX_1'\Phi^{-1} + X_1BX_2'\Phi^{-1} + X_2CX_1'^{-1}\Phi^{-1} + X_2DX_2'\Phi^{-1} \\
                &= H_1  - H_1H_2 - H_2H_1  + H_2 + H_2H_1H_2 \\
                &= (I - H_2)(H_1  - H_1H_2) + H_2\\
                &= (I - H_2)H + H_2 \\
                &= H + H_2(I - H).
\end{align*}

From Sherman-Morrison, $\tilde{\Phi} = (\Phi + X_2\Sigma X_2')^{-1}$ is positive-definite.  Then $H$ is a projection matrix onto the column space of $X_1$, with inner product $\tilde{\Phi}^{-1}$, and as $\mathbf{1} \in \text{span}(X_1)$, $H\mathbf{1} = \mathbf{1}$ and $(I - H)\mathbf{1} = 0$. The result follows.

\end{proof}

\subsection{Proof for Theorem~\ref{ssbf-thm:sfunder1}}\label{ssbf-appx:prfsfunder1}
We re-state Theorem~\ref{ssbf-thm:sfunder1} below for reference:

Theorem 2: Under the same setting as in Theorem~\ref{ssbf-thm:sumsto1}, let the shrinkage factor be defined as in (\ref{ssbf-eqn:sf}). Then given a point estimate $\hat{Y}_i$,  $0 < b_{iB_i} \le 1$ and likewise $0 <= b_{iL_i} < 1$, where $b_{iB_i}$ is the shrinkage factor and $b_{iL_i}$ the pooling factor.

The proof here is based on our earlier work, Lemma 1 in the supplementary material for \citet{zhang2020approximate}, and is re-created below for reference.

\begin{proof}

$b_{iB_i} > 0$: $b_{iB_i} = n_iw_{ii}.$ $V$ non-singular and $T$, $\Sigma$ positive-definite imply $V$ is positive-definite and $XVX'$ is positive semi-definite. Then the diagonal entries of $XVX'$ are non-negative and $w_{ii} = (XVX')_{ii}T_{ii}^{-1} > 0$.
\vspace{1em}

$b_{iB_i} \le 1$: Let $V_{-i}$ as in (\ref{ssbf-eqn:axe_linregr}), where the subscript $_{-i}$ indicates using the design matrix without the borrower cluster, $X_{-B_i}$, in place of $X$. We can solve for $V$ as a function of $V_{-i}$ using the Sherman-Morrison formula,
\begin{align}
    V &= (V_{-i} + x_ix_i'\phi_i^{-2})^{-1} \notag \\
    &= V_{-i} - \frac{n_i}{\phi_i^2}\frac{1}{1 + \frac{n_i}{\phi_i^2}x_i'V_{-i}x_i}V_{-i}x_ix_i'V_{-i}. \label{ssbf-itm:vsize}
\end{align}

As $V_{-i}$ is positive-definite and $\frac{n_i}{\phi_i^2}x_i'V_{-i}x_i \ge 0$, (\ref{ssbf-itm:vsize}) implies that $V_{-i} - V$ is positive semi-definite. Now, solving for $V_{-i}$ as a function of $V$ yields  
\begin{align}
    V_{-i} &= (V - x_ix_i'\phi_i^{-1})^{-1} \notag \\
     &= V + \frac{n_i}{\phi_i^2}\frac{1}{1 - \frac{n_i}{\phi_i^2}x_i'Vx_i}Vx_ix_i'V,
 \label{ssbf-itm:vjge}
\end{align}
through another application of Sherman-Morrison. As $({1 - \frac{n_i}{\phi_i^2})^{-1}x_i'Vx_i}Vx_ix_i'V$ is positive semi-definite and $b_{iB_i} = n_i\phi^{-2} x_i'Vx_i > 0$, $b_{iB_i}$ must be $\le 1$. 

\vspace{1em}
$0 <= b_{iL_i} < 1$: Theorem~\ref{ssbf-thm:sumsto1} and $0 < b_{iB_i} \le 1$ implies $0 <= b_{iL_i} < 1$.
\end{proof}


\section*{Appendix 2}
Appendix 2 illustrates the relationship between SSBF and influence analysis metrics RVSI and $S_i$, discussed in Section~\ref{ssbf-sec:thms_ssbf}. \ref{ssbf-appx:rvsi} derives the relationship to RVSI. \ref{ssbf-appx:si} derives the relationship to $S_i$.

\subsection{Relationship between RVSI and SSBF}\label{ssbf-appx:rvsi}

Value of information is an approach to outlier and influence analysis within the Bayesian literature that quantifies the value of sample information $Y_j$ using the reduction in loss that results from including $Y_j$ vs excluding it. For example, if $a_{Y_{-j}}$ is the estimator based on all data excluding $Y_j$ and $a_{Y_{-j}, Y_j}$ is the estimator for $Y_i$ based on all data, then the retrospective value of sample information (RVSI) under squared loss is
\begin{equation}\label{ssbf-eqn:rvsi_appx}
\text{RVSI}(Y_j | Y_{-j}; Y_i) = (a_{Y_{-j}} - a_{Y_{-j}, Y_j})'(a_{Y_{-j}} - a_{Y_{-j}, Y_j}).
\end{equation}
This can be explicitly written in terms of partial SSBF. Let response vector $\bm{Y}$ follow a normal linear regression with model design matrix $X$ as in (\ref{ssbf-eqn:normlinreg}) and let $\bm{Y}_j \in \RR^{n_j} \sim N(x_j'\bm{\beta}, \phi_j^2)$.

\citet{zhang2020approximate} showed that, for Bayesian hierarchical regression models, $E[x_j'\bm{\beta} | Y_{-j}, \hat{\Sigma}, \hat{\phi}] = E[x_j'\bm{\beta} | Y_{-j}](1 + O(P_2^{-1}))$, for posterior means $\hat{\Sigma}$ and $\hat{\phi}$. Taking as our estimators $a_{Y_{-j}} = E[x_i'\bm{\beta} | Y_{-j}, \hat{\Sigma}, \hat{\phi}]$ and $a_{Y_{-j}, Y_j} = E[x_i'\bm{\beta} | \bm{Y}, \hat{\Sigma}, \hat{\phi}]$ then approximates RVSI in (\ref{ssbf-eqn:rvsi_appx}) with $O(P_2^{-1})$ error.

Applications of the Sherman-Morrison formula and some algebra show that 
\begin{equation}
    E[\bm{\beta} | Y_{-j}, \hat{\Sigma}, \hat{\Phi}] = E[\bm{\beta} | \bm{Y}, \hat{\Sigma}, \hat{\Phi}] + \frac{n_j}{\phi_j^2}\frac{\hat{Y}_j - \bar{Y}_j}{1 - \frac{n_j}{\phi_j^2}x_j'Vx_j}Vx_j,
\end{equation}
and the difference in our estimators can then be written as the product of the average residual for $Y_j$ and their borrowing factor $n_jw_{ij}$,
\begin{equation}\label{ssbf-eqn:rvsi_a}
a_{Y_{-j}} - a_{Y_{-j}, Y_j} =\frac{w_{ij}}{b_{jL_j}}\frac{n_j(\hat{Y}_j - \bar{Y}_j)}{\phi^2},
\end{equation}
where $b_{jL_j}$ denotes the pooling factor for $\hat{Y}_j$.

Combining (\ref{ssbf-eqn:rvsi_appx}) and (\ref{ssbf-eqn:rvsi_a}), RVSI can be written as the product of the sum of squared residuals and PSSBF,
\begin{equation*}
   \text{RVSI}(Y_j | Y_{-j}; Y_i) = \frac{\text{PSSBF}_{ij}}{b_{jL_j}^2}  \frac{n_j(\hat{Y}_j - \bar{Y}_j)^2}{\phi^4}(1 + O(P_2^{-1}).
\end{equation*}

\subsection{Relationship between $S_i$ and SSBF}\label{ssbf-appx:si}
\citeauthor{pena2005new}'s $S_i$ is the squared norm of the standardized vector $\bm{s}_i = (\hat{Y}_i - \hat{Y}_{i(1)}, \dots, \hat{Y}_i - \hat{Y}_{i(N)})'$, where $\hat{Y}_{i(j)}=E[Y_i | Y_{-j}]$. $S_i$ can be re-written as a linear combination of Cook's distances, $D_j$,
\begin{equation*}
    S_i = \frac{\bm{s_i}'\bm{s}_i}{p\hat{var}(\hat{Y}_i)} = \sum_{n = 1}^N \frac{w_{in}^2}{w_{ii}w_{nn}}D_n, \quad D_n = \frac{e_n^2}{ps^2}\frac{w_{nn}}{(1 - w_{nn})^2}
\end{equation*}
where $D_n$ is the Cook's distance for ${Y}_n$, $\bm{e} = \bm{Y} - X\bm{\hat{\bm{\beta}}}$, ${e}_n = (Y_n - x_n'\bm{\hat{\beta}})$, and $s^2 = \bm{e}'\bm{e}/(n -P)$, where $P$ is the dimension of $\bm{\beta}$.

If $\bm{Y}_j \in \RR^{n_j} \sim N(x_j'\bm{\beta}, \phi_j^2)$, then $w_{ik} = w_{ik'}$ for all $i \in \{1, \dots, N\}$ and all $k, k' \in j$, and we can aggregate over the clusters of data $\bm{Y}_j$ to obtain  
\begin{equation*}
    S_i = \sum_{j}\frac{\text{PSSBF}_{ij}}{w_{ii}w_{jj}}\bar{D}_j, \quad \bar{D}_j = \frac{\bar{e}_j^2}{ps^2}\frac{w_{jj}}{(1 - w_{jj})^2}.
\end{equation*}

\section*{Appendix 3}

\subsection{Borrowing factors for the Radon example}\label{ssbf-appx:bal}

For $b_{k'j}$ to be the borrowing factor for the contrast in data means $\bar{Y}_{k'j} - \bar{Y}_{k'j'}$, it is necessary to show that for all lenders $g, g'$ corresponding to $\bm{Y}_{k'j'}$, $x_i'Vx_g = x_i'Vx_{g'}$.   

For the model in (\ref{ssbf-eqn:radonsetup}), let $N$ denote the dimension of $\bm{Y}$. Under a balanced data scenario, the number of houses in any county $j$ with any basement status $k$ is $n :=  N/(2J)$. As we are conditioning on the continuous covariate $u_j$, we note that 
\begin{equation}\label{ssbf-eqn:vinvbal}
    V^{-1} = \begin{bmatrix}
       \frac{N}{2\phi^2} & 0 & \frac{n}{\phi^2} \bm{1}_J' \\
       0 & \frac{N}{2\phi^2} & \frac{n}{\phi^2} \bm{1}_J' \\
       \frac{n}{\phi^2} \bm{1}_J & \frac{n}{\phi^2} \bm{1}_J & \left(\frac{n}{\phi^2} + \frac{1}{\sigma^2}\right)I_J
    \end{bmatrix},
\end{equation}
where $\bm{1}_J \in \RR^J$ is the vector of ones and $I_J \in \RR^{J \times J}$ is the identity matrix. 

For $Y_g$ and $Y_{g'}$ within the same relationship group (e.g., same-county lenders, same-basement lenders, or others), the only difference between $x_g$ and $x_{g'}$ is the indicator variable for the county-specific effect. Now let $P = J + 2$, the number of columns in $V$ and let $M_{gg'} \in \RR^{P \times P}$ be the permutation matrix such that $M_{gg'}x_{g'} = x_g$. Then $\tilde{X} \coloneqq XM_{gg'}$ is the model design matrix with the columns corresponding to indicator variables for counties $g$ and $g'$ switched. As the data are balanced, using $\tilde{X}$ instead of $X$ still results in (\ref{ssbf-eqn:vinvbal}) and so
\begin{equation}
    V = \left(\tilde{X}'\tilde{X}/\phi^2 + \begin{bmatrix}0 & 0 \\ 0 & \frac{1}{\sigma^2}I\end{bmatrix}\right)^{-1} = \left({X}'{X}/\phi^2 + \begin{bmatrix}0 & 0 \\ 0 & \frac{1}{\sigma^2}I\end{bmatrix}\right)^{-1}.
\end{equation}

This implies that 
\begin{equation}
   Vx_{g'} = VM_{gg'}x_{g'} = Vx_g.
\end{equation}

Since $x_i'Vx_g = x_i'Vx_{g'}$ for all lenders $g, g'$ in the same relationship group, we can formulate the point estimate $\hat{u}_{kj}$ as a weighted sum of relationship group means,
\begin{equation}
    \hat{\mu}_{kj} = b_{kj}\bar{Y}_{kj}  + b_{kj'}\bar{Y}_{kj'} + b_{k'j}(\bar{Y}_{k'j} - \bar{Y}_{k'j'},
\end{equation}
where $b_{kj}$ is the shrinkage factor and $b_{kj'}$ is the pooling factor. When $J$ is large, this contrast in means, given $\bm{\beta}$ and $a$, has expected value of $a_j$. Then $b_{k'j} = -b_{k'j'}$ isolates the county-specific effect $a_j$ and represents borrowing from lenders due to $a_j$. Similarly, $b_{kj'}$ represents borrowing due to the basement intercept.

\subsection{Supplemental figures for Scottish respiratory disease example}\label{ssbf-appx:srd-figs}

\begin{figure}[!ht]
    \centering
    \includegraphics[width=0.4\linewidth]{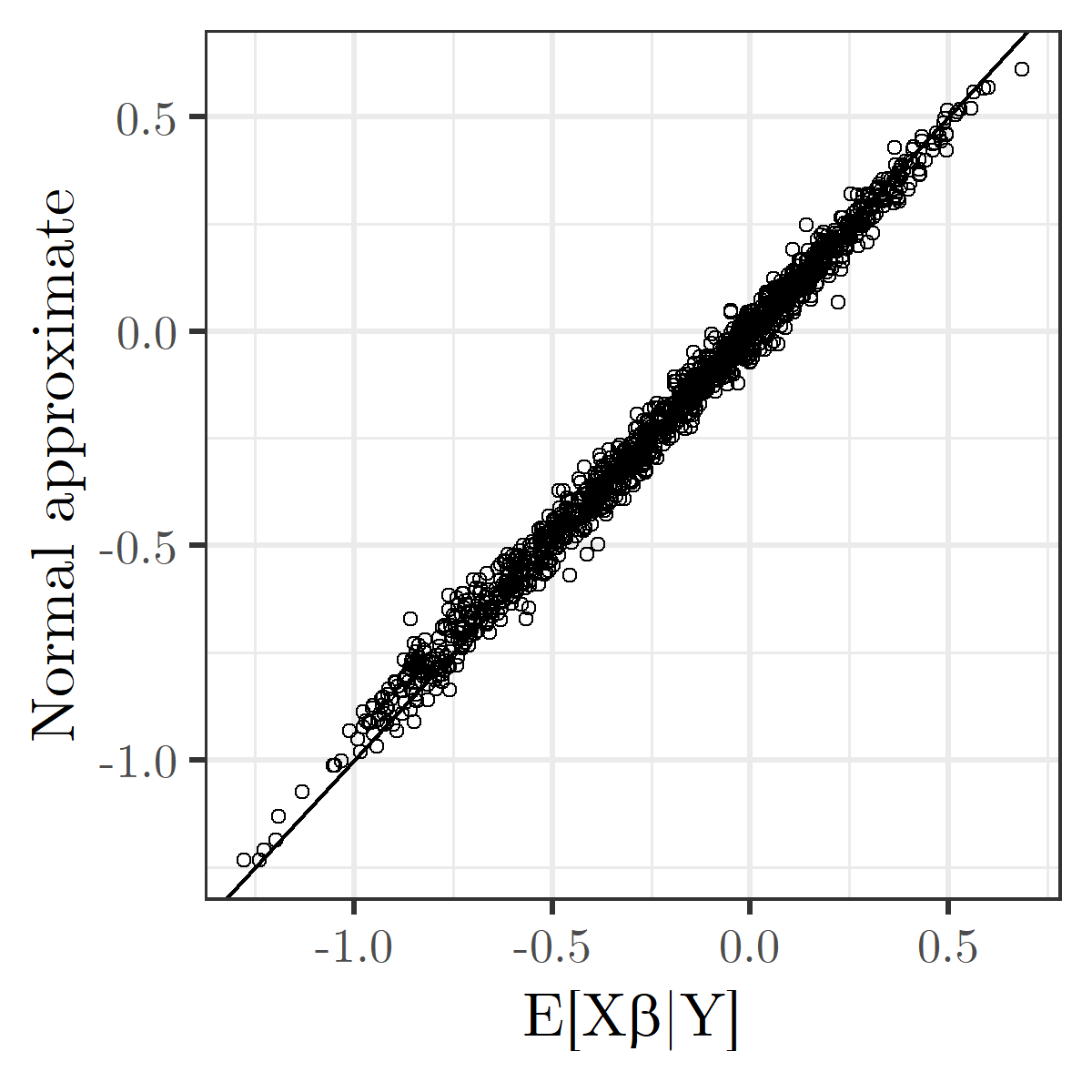}
    \caption{Scatterplot of point estimates obtained through normal approximation (y-axis) versus actual posterior means $E[X\beta | Y]$ (x-axis) for the Scottish respiratory disease data. The normal approximation used is (\ref{ssbf-eqn:srd-normapprox}).}
    \label{ssbf-fig:normapprox}
\end{figure}

\begin{figure}[!ht]
    \centering
    \includegraphics[height=12em]{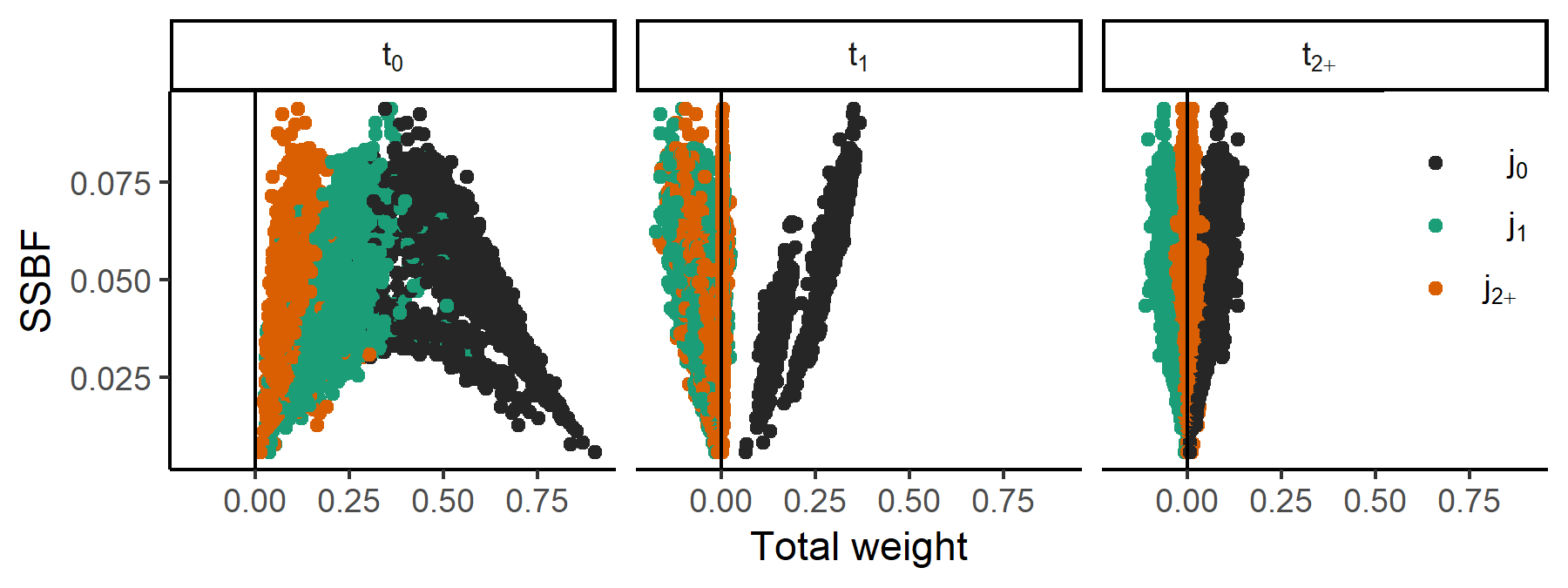}
    \caption{A scatter plot of SSBF against the total weight applied to lender relationship groups, where each panel represents a different temporal relationship group ($t_0$, $t_1$, $t_{2+}$ for same year, adjacent year, other years, respectively) and colors represent different spatial relationship groups (black for $j_0$, green for $j_1$, orange for $j_{2+}$, corresponding to same IG, neighboring IG, and farther IGs, respectively).}
    \label{ssbf-fig:srd-totwt}
\end{figure}

\begin{figure}[!ht]
    \centering
    \includegraphics[width=0.8\linewidth]{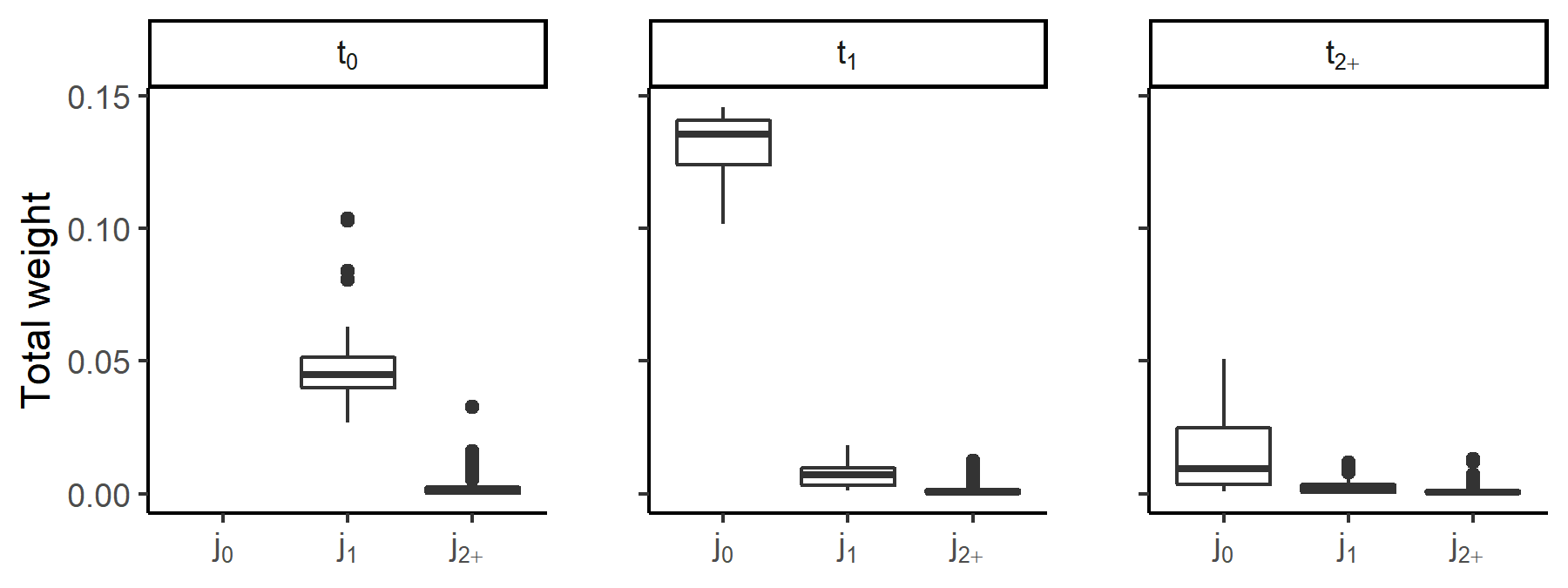}
    \caption{Boxplots of total (absolute) weight placed on 11 influential points when $\hat{\alpha} = 0.57$ and $\hat{\rho} = 0.76$. Box plots are split into temporal ($t_0$, $t_1$, $t_{2+}$) and spatial ($j_0$, $j_1$, $j_{2+}$) relationship groups. The plots do not include the shrinkage factor, hence no boxplot for $b_{t_0j_0}$.}
    \label{ssbf-fig:inflalt}
\end{figure}


\bibliographystyle{abbrvnat}  
\bibliography{ref}

\end{document}